\documentclass[twocolumn,prl,amsmath,amssymb,showpacs,superscriptaddress,floatfix]{revtex4}
\usepackage{graphicx}% Include figure files
\usepackage{bm}% bold math
\usepackage{lineno,hyperref}

\usepackage{epsf}
\usepackage{epstopdf}
\begin{document}
%\draft
\title{Melting scenarios of two-dimensional Hertzian spheres with a single triangular lattice}%:
%First-Order versus Continuous Transition }

%\author{E. N. Tsiok, Yu. D. Fomin, E. A. Gaiduk, and V. N. Ryzhov}
%\affiliation{ Institute for High Pressure Physics RAS, 108840
%Kaluzhskoe shosse, 14, Troitsk, Moscow, Russia}

\author{E. N. Tsiok}
\affiliation{Institute for High Pressure Physics, Russian Academy
of Sciences, Troitsk 108840, Moscow, Russia}

\author{E. A. Gaiduk}
\affiliation{Institute for High Pressure Physics, Russian Academy
of Sciences, Troitsk 108840, Moscow, Russia}

\author{Yu. D. Fomin}
\affiliation{Institute for High Pressure Physics, Russian Academy
of Sciences, Troitsk 108840, Moscow, Russia}
\affiliation{Moscow Institute of Physics and Technology, 9 Institutskiy Lane, Dolgoprudny City, Moscow Region, Russia }

\author{V. N. Ryzhov}
\affiliation{Institute for High Pressure Physics, Russian Academy
of Sciences, Troitsk 108840, Moscow, Russia}

\date{\today}

\begin{abstract}
We present a molecular dynamics simulation study of the phase
diagram and melting scenarios of two-dimensional Hertzian spheres
with exponent 7/2. We have found multiple re-entrant melting of a
single crystal with a triangular lattice in a wide range of
densities from 0.5 to 10.0. Depending on the position on the phase
diagram, the triangular crystal has been shown to melt through
both two-stage melting with a first-order hexatic - isotropic
liquid transition and a continuous solid - hexatic transition as
well as in accordance with the
Berezinskii-Kosterlitz-Thouless-Halperin-Nelson-Young (BKTHNY)
scenario (two continuous transitions with an intermediate hexatic
phase). We studied the behavior of heat capacity and have shown
that despite two-stage melting, the heat capacity has one peak
which seems to correspond to a solid-hexatic transition.
\end{abstract}


\maketitle

\section{Introduction}

The behavior of soft/deformable colloidal mesoparticle systems
such as dendrimers, star polymers, and block-copolymer micelles
presents much interest for interdisciplinary studies in physical
chemistry, materials science, and soft matter \cite{likos,c1,c2}.
Today, the question of how to link the  interaction potential
shape to phase behavior  still remains one of the serious
challenges in the field of soft matter and is very important for
the development of materials with novel optical, mechanical, and
electronic properties \cite{c3,c4,c44,c5,c6,c7,c8,c9,c10,c11}. At
high enough densities, polymeric nanocolloids self-organize in a
number of ordered structures including close-packed and
nonclose-packed crystalline phases \cite{c12,c13,c14,c15}. The
naive point of view is that the variety of complex crystalline
phases is a result of colloid shape anisotropy and/or directional
(anisotropic) interactions, however, it was shown that different
non-close-packed structures could be obtained using  isotropic
potentials like, for instance,  two scale repulsive shoulder
potentials \cite{rice,dfrt1,dfrt2,dfrt3,dfrt5,we5,rice_jcp}. At
the same time, there is a great variety of nontrivial
phenomenological interactions, some of which even lead to complete
overlap among the components and demonstrate very rich phase
behavior \cite{likos,c4,c13,c14,c15}. A popular possibility is to
view particles as elastic spheres which repel each other on
contact (Hertzian spheres). For small deformations, repulsion is
additive-pair-wise and according to the Hertz theory proportional
to $h^{5/2}$, where $h$ is the indentation of the contact zone
\cite{c4}. In three dimensions this power law gives rise to a
complex phase diagram, including water-like anomalous behavior
\cite{hertz3d,rosbreak}. Recently, the behavior of two-dimensional
Hertzian spheres has become a topic of considerable interest
\cite{molphys,miller,hertzmelt,hertzqc}.

The generalized Hertzian potential used for the phenomenological
description of  deformable colloidal particles has the following
form:
\begin{equation}\label{pot}
   U(r)=\varepsilon \left ( 1- r/ \sigma \right)^{\alpha}H(1-r),
\end{equation}
where $H(r)$ is the Heaviside step function and parameters
$\varepsilon$ and $\sigma$ set the energy and length scales. The
Hertzian potential with $\alpha$=5/2 corresponds to the elastic
energy of deformation of two spheres.

%Investigation of two-dimensional ($2D$) and quasi-$2D$ systems has been a perennial research interest, because such systems demonstrate the generic features of phase behaviors of %many functional materials including
%graphene, protein membranes, nanocrystals on graphite, etc. For instance, $2D$ colloidal crystals can be used as structured substrates and seeds for bulk photonic crystals employed %for sensors and light conversions, in optical composites, and for spectroscopy using optical field localization (see Refs. ).

Due to well-developed fluctuations and influence of confinement,
2D systems demonstrate many unusual features which cannot be seen
in three dimensional systems with the same type of inter-particle
interaction. The most striking example is the melting of 2D
crystals. While in three dimensions melting occurs as a first
order phase transition only, it seems that there are three
different melting scenarios of 2D crystals which are the most
popular at the moment (see, for instance, review \cite{pu2017} and
the references therein). As in the 3D case, melting can occur
through one first-order transition
\cite{chui83,ryzhovJETP,RT1,rto1,rto2}. However, in addition,
there are two completely different scenarios. The first one is the
Berezinskii-Kosterlitz-Thouless-Halperin-Nelson-Young (BKTHNY)
theory which is widely accepted now. This theory has support in
computer simulations and real experiments
\cite{pu2017,ber1,ber2,kosthoul72,kosthoul73,kost,halpnel1,halpnel2,halpnel3,str,keim1,zanh,keim2,keim3,keim4}).
According to this theory 2D solids melt through two
Berezinskii-Kosterlitz-Thouless (BKT) type transitions
\cite{ber1,ber2,kosthoul72,kosthoul73,kost}. At the first
transition the dissociation of bound dislocation pairs occurs,
which leads to the transformation of long-range orientational
order into quasi-long-range one, and quasi-long-range positional
order becomes short-range. The new intermediate phase with
quasi-long-range orientational order is called a hexatic phase
with a zero shear modulus. The hexatic phase can be considered as
a quasi-ordered liquid. In the course of the second transition the
hexatic phase transforms into an isotropic liquid with short-range
orientational and positional orders through unbinding disclination
pairs. It should be noted, that strictly speaking the BKTHNY
theory only exists for triangular lattices. There are no
renormalization group equations for lattices with other
symmetries. The BKTHNY theory only provides the limits of
stability of solid and hexatic phases.

Another melting scenario was proposed in Refs.
\cite{foh1,foh2,foh3,foh4,foh6}. In computer simulations
\cite{foh1,foh2,foh3,foh4,foh6} and in experiments \cite{hsn} it
was shown that the system could melt through a continuous BKT type
solid-hexatic transition, but the hexatic-liquid transition is of
the first order. In paper \cite{foh4} (see also \cite{wemp,dfrt6})
a detailed study of a soft disk system with potential
$U(r)=(\sigma/r)^n$ was presented. It was shown that the system
melted in accordance with the BKTHNY scenario for $n \leq 6$, while
for $n>6$ a two-stage melting transition took place with a
continuous solid-hexatic and first-order hexatic-liquid
transition.

For a long time only close-packed triangular crystal structures
were observed in experimental studies, and the BKTHNY theory was
good for analysis of melting of these systems. However, in the
last decades many experimental investigations and computer
simulations have demonstrated that 2D and quasi-2D systems also
demonstrate other crystalline structures. Square ice formation
when water is confined between two graphene planes was reported in
Ref. \cite{geim}. The square phase of a one atom thick iron layer
in graphene was also observed in Ref. \cite{iron}. Even more
complex phases were observed in 2D colloidal systems
\cite{dobnikar}. However, until now experimental observations of
ordered 2D structures other than a triangular lattice are rather
rare.

At the same time many computational works report the appearance of
complex 2D structures including a square phase
\cite{dfrt1,dfrt2,dfrt3,jain,marcotte}, a honeycomb lattice
\cite{jain,marcotte}, a Kagome lattice \cite{pineros}, different
quasicrystalline phases \cite{qc1,qc2,qc3,we5}, etc.

Hertzian spheres demonstrate extremely complex phase diagrams and
liquid state anomalies in both three-dimensional and
two-dimensional spaces (see Refs. \cite{hertz3d} and
\cite{rosbreak} for a phase diagram and anomalous behavior of 3D
Hertzian spheres and \cite{molphys} for the 2D case).

Two-dimensional Hertzian spheres demonstrate a large variety of
different ordered phases, including a dodecagonal quasicrystal
\cite{molphys}. Interestingly, Hertzian spheres with $\alpha$=5/2
demonstrate all three melting scenarios in the same system and
tricritical points where a change of the melting scenarios occurs
\cite{molphys,hertzmelt}).

\begin{figure}
\includegraphics[width=8cm]{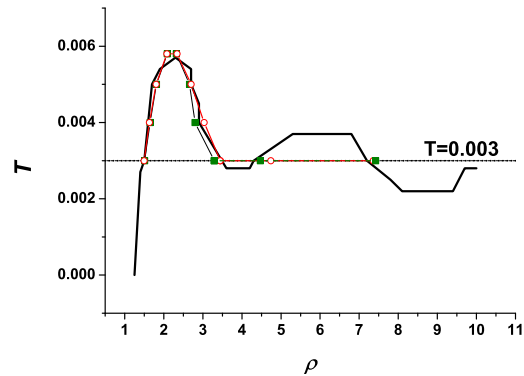}%

\caption{\label{pd} The phase diagram of a Hertz system with
$\alpha$=7/2. The line shows the data from \cite{miller}. The
symbols show our results (see the text). The horizontal line marks
the $T$=0.003 isotherm at which the study of the phase diagram at
high density is performed in the present paper.}
\end{figure}

The phase diagram of the system is also extremely sensitive to
control parameter $\alpha$. In Ref. \cite{miller} three different
values of $\alpha$ were considered: 3/2, 5/2 and 7/2. It was found
that increasing $\alpha$ led to a decrease in the number of
different phases in the system. This result was confirmed in Ref.
\cite{hertzqc} where a very elaborate study of phases appearing in
a Hertzian system with $\alpha$ ranging from 2 to 3 was performed.
It was shown that at the values of $\alpha < 2.2$ there were many
different ordered phases. In particular, complex phases like a
sigma phase or a kite phase were observed in the system. Moreover,
a Hertz system with $\alpha<2.75$ could demonstrate
quasicrystalline phases \cite{molphys,hertzqc}. However, at
$\alpha$=3 only triangular and rhombic phases were found
\cite{hertzqc}. In Ref. \cite{miller} a Hertzian system with
$\alpha$=7/2 was studied and it was shown that the only
crystalline phase appearing in the system was a triangular crystal
(Fig.~\ref{pd}). At the same time the melting line of the system
appears to be very complex. In spite of the single crystalline
phase in the system the melting line is non-monotonous, with
several maxima and minima. Such a complex shape of the single
phase melting line is extremely interesting and deserves to be
studied. Moreover, as it was shown in recent publications
\cite{molphys,hertzmelt},the triangular phase of a Hertzian system
with $\alpha$=5/2 melted through different scenarios at different
densities. One can expect that Hertzian spheres with $\alpha$=7/2
also demonstrate different melting scenarios.

Another intriguing problem that is rarely discussed in literature
is investigating heat capacity behavior in 2D melting. The
peculiarities of heat capacity behavior in 2D melting can be
related to a change in the density of topological defects. The BKT
transition is an infinite order (continuous) phase transition.
Continuous phase transitions are accompanied by divergences in
thermodynamic quantities caused by the divergence of the
correlation length as critical temperature $T_c$ is approached.
For the BKT transition there is an exponential divergence of
isochoric heat capacity $c_v(T)$ \cite{ch_l}. It is so rapid and
occurs within such a narrow temperature range that the divergence
in $c_V$ cannot be resolved in simulations or in experiments
\cite{ch_l}. The value of $c_v$ measured for the BKT transition in
an $X-Y$ model is sketched in Fig. 9.4.3 in Ref. \cite{ch_l} and
has unobservable essential singularity at $T_c$ and and a
nonuniversal model-dependent maximum above $T_c$ associated with
the entropy liberated by the unbinding of bound topological defect
pairs. Using large-scale Monte Carlo simulation for the BKTHNY
transition in a Lennard-Jones system \cite{wierschem} the authors
observed one broad peak in the specific heat per particle. In Ref.
\cite{cv2d} a melting transition of superparamagnetic colloidal
spheres confined in two dimensions was studied. This system had
long-range interaction and melted in accordance with the BKTHNY
scenario, and it was shown that there was only one smooth peak
inside the hexatic phase. At the same time in non-ideal Yukawa
systems \cite{vaulina} the BKTHNY transition is accompanied by two
singularities ("jumps") for the heat capacities near temperatures
of the fluid-to-hexatic phase and hexatic-to-"perfect" (without
defects) crystal transitions. The question is whether in the case
of two-stage melting one would observe two peaks on the $c_v(T)$
curve and where these possible peaks would be located or there is
only one peak related to the melting transition.

The goal of the present paper is to study the melting scenarios of
a 2D Hertzian sphere system with control parameter $\alpha$=7/2
and to investigate the behavior of heat capacity $c_v(T)$ across
the melting line.

\section{System and methods}

The simulation setup is similar to the one described in Ref.
\cite{molphys}. We use rescaled number density
$\tilde{\rho}=N\sigma^2/A$ and temperature $\tilde{T}=k_B
T/\varepsilon$ and omit tildes in what follows. Firstly, we
simulated a small system of 5000 particles in a rectangular box
under periodic boundary conditions in a wide range of densities
from $\rho_{min}$=0.5 to $\rho_{max}$=10.0 at temperature
$T$=0.003. The molecular dynamics in a canonical ensemble
(constant number of particles $N$, square of the system $A$, and
temperature $T$) was used. According to Ref. \cite{miller}, at
this temperature there are two crystalline regions on this way:
the first one in the range of densities (approximately) 1.5 - 3.5
and the second one in range 4.4 - 7.2. We roughly localize phase
boundaries from the equations of state and radial distribution
functions (rdfs) in a small system. The small systems are
propagated for $5 \cdot 10^6$ steps with time step $dt$=0.001. The
first $2.5 \cdot 10^6$ steps were used for equilibration and the
last $2.5 \cdot 10^6$ were used for production.

At the second step we used larger systems and smaller density
intervals in order to locate  precise phase boundaries. We used a
system of 20000 particles at densities below 3.6 and 45000
particles above this threshold. The number of particles was
increased at higher densities in order to evaluate the spatial
correlation functions at large enough separations.

Special attention was paid to the region of the first maximum on
the phase diagram. In this case we calculated a complete melting
line up to a maximum. The simulation methodology was the same: we
roughly localized phase boundaries simulating a small system and
after that we performed simulations with 20000 particle systems to
find precise phase boundaries.

In order to estimate the phase boundaries we used the method based
on a combination of equations of state, orientational and
translational order parameters and their correlation functions.
The local orientational order parameter (OOP) was defined in the
following way \cite{halpnel1,halpnel2,dfrt5,dfrt6}:

\begin{equation}
\psi_6({\bf r_i})=\frac{1}{n(i)}\sum_{j=1}^{n(i)} e^{i
n\theta_{ij}}\label{psi6loc},
\end{equation}
where $\theta_{ij}$ is the angle of the vector between particles
$i$ and $j$ with respect to the reference axis and the sum over
$j$ is counting the $n(i)$ nearest-neighbors of $j$, obtained from
the Voronoi construction. The global OOP can be calculated as an
average over all particles:
\begin{equation}
\Psi_6=\frac{1}{N}\left<\left|\sum_i \psi_6({\bf
r}_i)\right|\right>.\label{psi6}
\end{equation}

The translational order parameter (TOP) has the following form
\cite{halpnel1, halpnel2, dfrt5, dfrt6}:
\begin{equation}
\Psi_T=\frac{1}{N}\left<\left|\sum_i e^{i{\bf G
r}_i}\right|\right>, \label{psit}
\end{equation}
where ${\bf r}_i$ is the position vector of particle $i$ and {\bf
G} is the reciprocal-lattice vector of the first shell of the
crystal lattice.

The orientational correlation function (OCF) is defined as

\begin{equation}
g_6(r)=\frac{\left<\Psi_6({\bf r})\Psi_6^*({\bf 0})\right>}{g(r)},
\label{g6}
\end{equation}
where $g(r)=<\delta({\bf r}_i)\delta({\bf r}_j)>$  is the pair
distribution function. In the hexatic phase the long-range
behavior of $g_6(r)$ has the form $g_6(r)\propto r^{-\eta_6}$ with
$\eta_6 \leq \frac{1}{4}$ \cite{halpnel1, halpnel2}.

Analogously, the translational correlation function (TCF) is given
by
\begin{equation}
g _T(r)=\frac{<\exp(i{\bf G}({\bf r}_i-{\bf r}_j))>}{g(r)},
\label{GT}
\end{equation}
where $r=|{\bf r}_i-{\bf r}_j|$. In the solid phase the long-range
behavior of $G_T(r)$ has the form $g_T(r)\propto r^{-\eta_T}$ with
$\eta_T \leq \frac{1}{3}$ \cite{halpnel1, halpnel2}. In the
hexatic phase and isotropic liquid $g_T$ decays exponentially.

In addition, we also studied the behavior of isochoric heat
capacity along the phase transitions. The heat capacity was
calculated from the fluctuation formula
$c_V=\frac{<U^2>-<U>^2}{Nk_B(T)^2}$, where $U$ is the energy of
the system \cite{book_fs}. In the present paper we calculate the fluctuation
of potential energy only, and therefore the values of the heat capacity are the
difference between the full heat capacity and the ideal gas value $c_{V,id}=1$.
For the sake of brevity we call it simply as $c_V$.

At the second stage of the work we performed longer simulations of
$5 \cdot 10^7$ steps. Moreover, $2D$ systems demonstrate extremely
strong fluctuations, and the results were still noisy. In order to
remove the noise we simulated several replicas of the system. The
initial configuration was always a triangular crystal, but the
initial velocities were different. Up to 80 replicas were used;
however, in most cases 50 replicas gave acceptable accuracy. The
final pressures and heat capacities were calculated by averaging
over all replicas.

\section{Results and discussion}

\begin{figure}

\includegraphics[width=8cm]{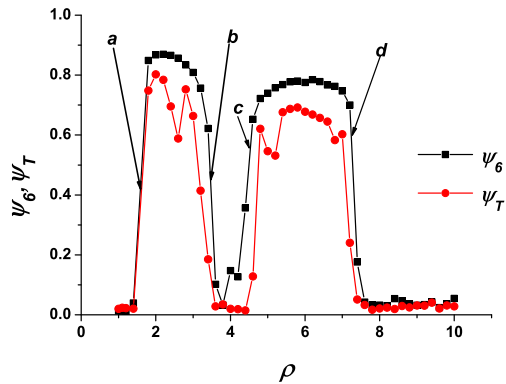}%

\caption{\label{small} Orientational and translational order
parameters $\psi_6$ and $\psi_T$ along isotherm $T$=0.003.}
\end{figure}

Firstly we studied the order parameters of a small system of 5000
particles (Fig. \ref{small}) at  temperature $T$=0.003. We
performed very quick calculations to roughly estimate phase
boundaries. Fig. \ref{small} shows the OOP and TOP of the system.
One can see that there are two regions where both order parameters
have finite values, i.e. the system has two zones where the
crystalline phase is stable. These regions are $\rho=1.4 \div 3.8$
and $\rho= 4.4 \div 7.8$. Therefore four regions of phase
transitions can be identified. In Fig. \ref{small} these regions
are marked with letters from $a$ to $d$. Below we will use these
letters to denote the regions of the phase diagram. In the
vicinity of regions $a$ and $c$, an increase in density leads to
the transformation of a liquid into solid, while at transition
borders $b$ and $d$ re-entrant melting of the solid takes place.
In order to obtain the phase boundaries more accurately, we
studied these regions with a larger system and performed longer
simulations (see the methods above).

It is common knowledge that 2D systems demonstrate strong
fluctuations. As a result one needs to collect a very large
statistic in order to get reliable results. Although we performed
rather long simulations ($5 \cdot 10^7$ steps) we found that the
accuracy of the results was not sufficient. Because of this we
performed more simulations with different initial velocities. We
called such configurations different replicas. Up to 80 replicas
of the same system were considered. Fig. \ref{pav} (a) shows the
equation of state of the system in region a for a different number
of replicas. One can see that the results for a single replica are
very noisy. In the case of 20 replicas the results are better, but
still not very accurate. We find that only averaging over 80
replicas gives acceptable accuracy.

\begin{figure}\label{pav}
\includegraphics[width=8cm]{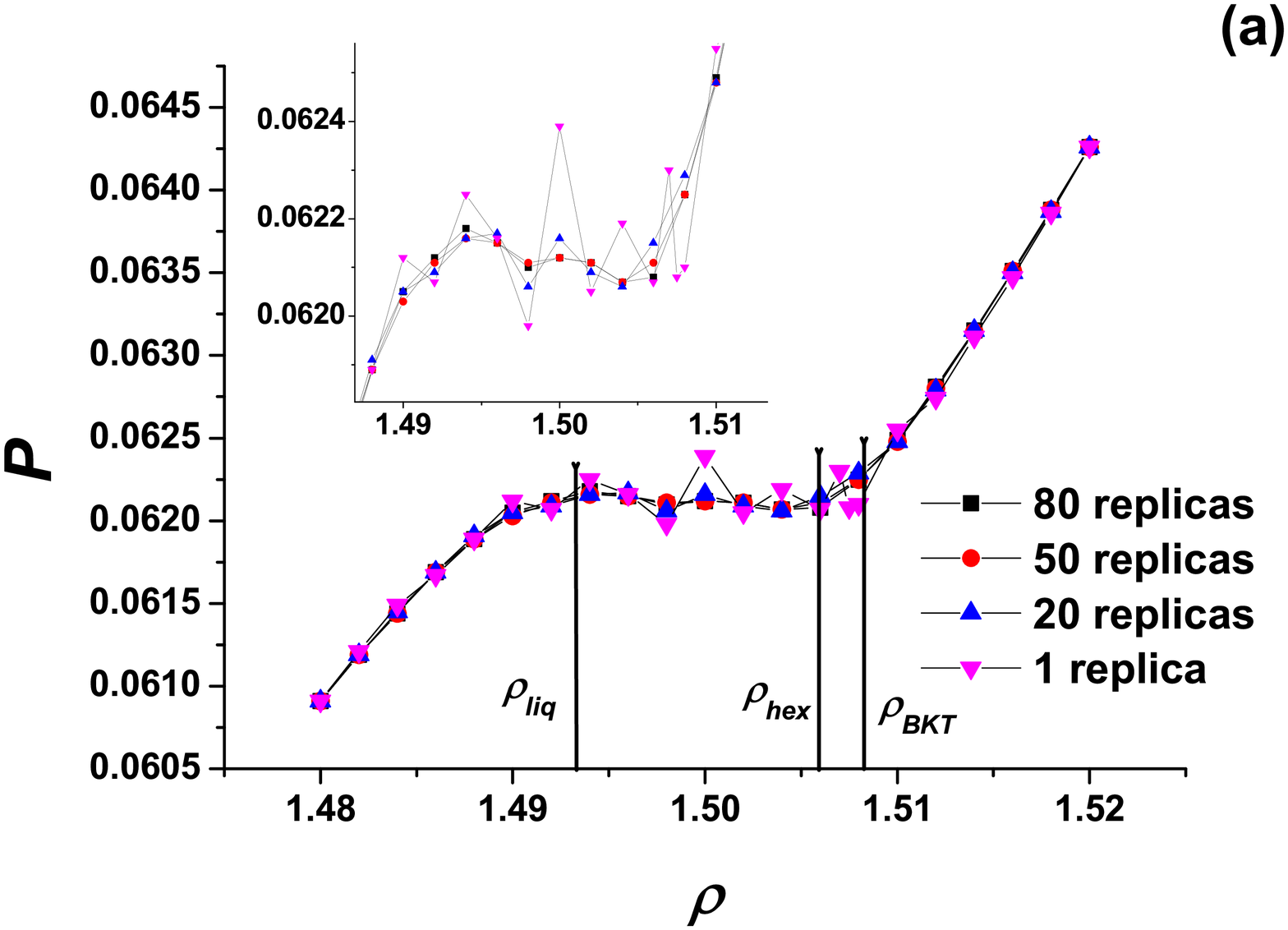}%

\includegraphics[width=8cm]{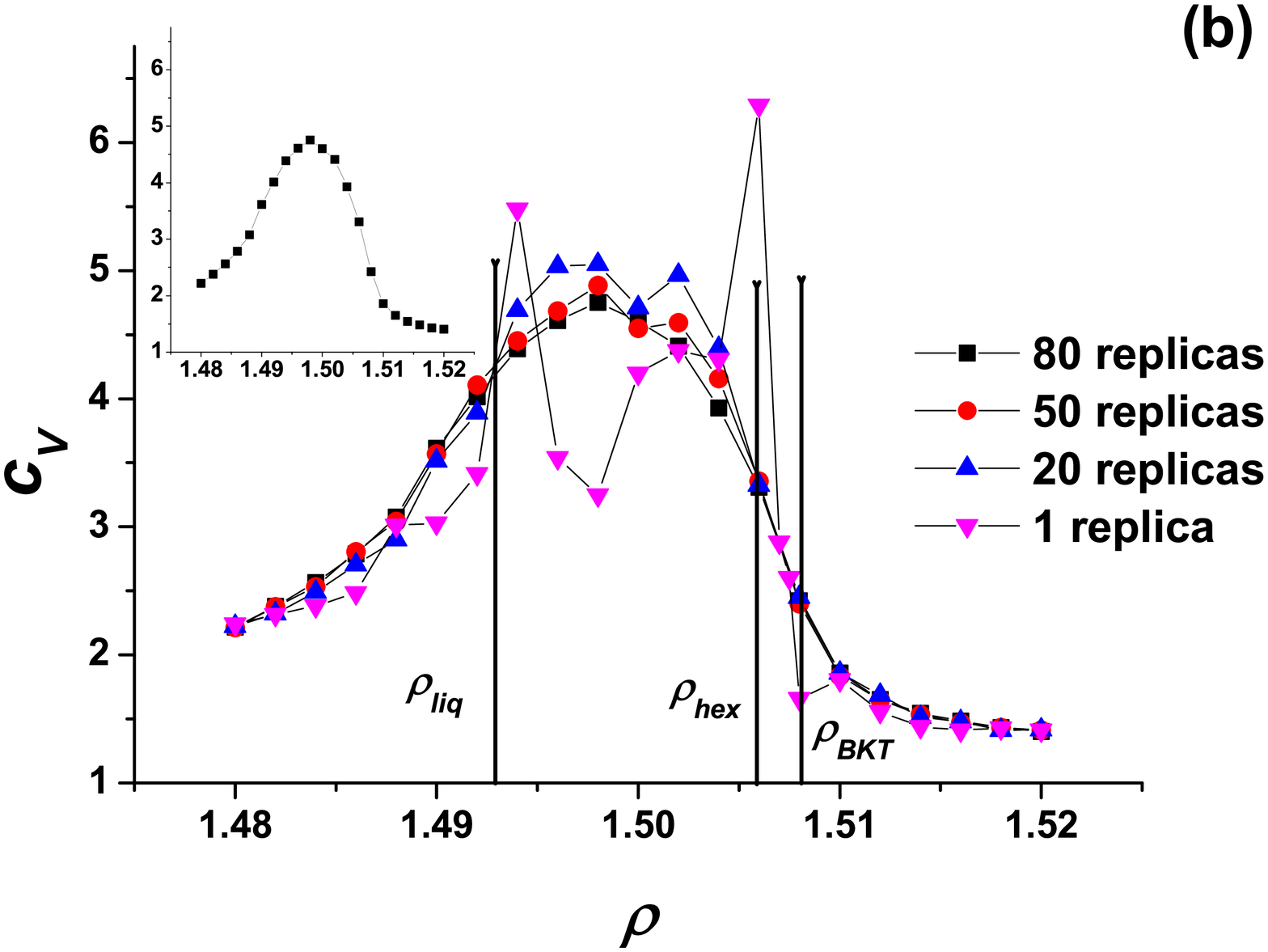}%

\caption{\label{pav} (a) Comparison of the equations of state for
80, 50, 20 and one replica of the system at $T$=0.003. The inset
enlarges the region of the Mayer-Wood loop. (b) Comparison of heat
capacity for 80, 50, 20 and one replica of the system at
$T$=0.003. The inset shows the heat capacity for 80 replicas.}
\end{figure}

\begin{figure}
\includegraphics[width=8cm]{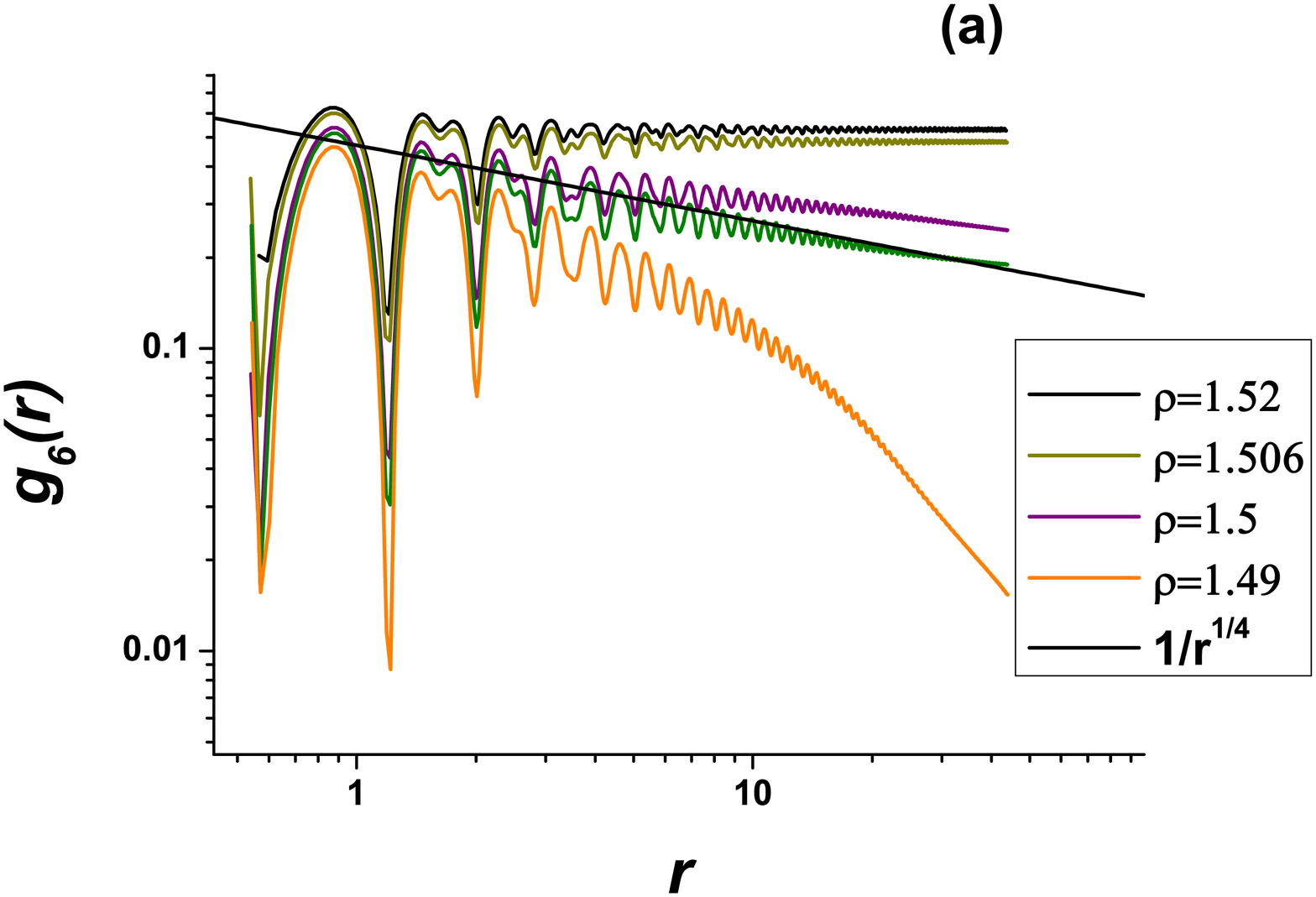}%

\includegraphics[width=8cm]{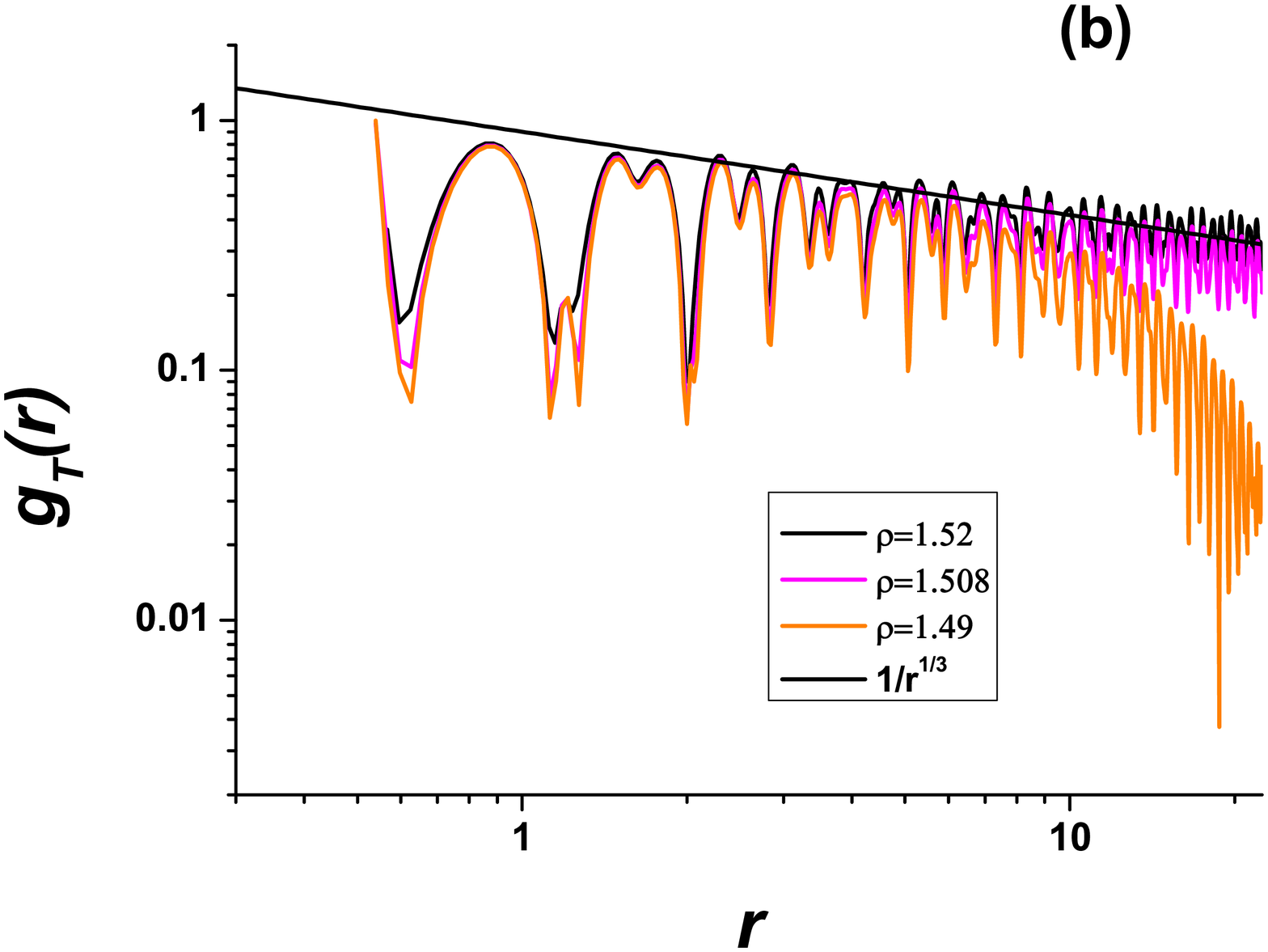}%

\caption{\label{psi-a} (a) The translational and (b) orientational
correlation functions of the system in the density region
corresponding to a transition from liquid into a triangular phase
at low density (region a in terms of Fig. \ref{small} (b)).
$T$=0.003.}
\end{figure}

After averaging over 80 replicas we observed that the equation of
state demonstrated the Mayer-Wood loop which signalizes the
presence of a first-order phase transition (Fig. \ref{pav} (a)).
In order to distinguish between the first and the third scenarios
of melting we studied  order parameters $\psi_6$ and $\psi_T$ and
their correlation functions in a system of 20000 particles. Fig.
\ref{psi-a} shows the translational and orientational correlation
functions of the system in the region of melting at the lowest
densities. One can see that the transition from the crystal into
hexatic phase occurs at density $\rho_{BKT}$=1.508. The
instability of the hexatic phase takes place at $\rho$=1.496,
which is in the middle of the Mayer-Wood loop. This situation
corresponds to the third melting scenario, i.e. the continuous BKT
transition from the solid to hexatic phase and the first-order one
from the hexatic phase to the isotropic liquid. The densities of
the hexatic and liquid phases at coexistence are
$\rho_{hex}$=1.506 and $\rho_{liq}$=1.493.

We also studied the isochoric heat capacity of the system (Fig.
\ref{pav} (b)). One can see that the heat capacity is more
sensitive to the quality of statistical averaging than the
equation of state. It demonstrates a single peak at $\rho$=1.498
after averaging over 80 replicas. Besides, $c_V$ demonstrates a
small second peak at $\rho$=1.502 with 50 replicas. One can
conclude that inappropriate averaging can lead to qualitatively
wrong results in the case of 2D systems. The real peak appears
inside the Mayer-Wood loop of the hexatic to liquid transition, in
the neighborhood of the hexatic phase stability limit. As
discussed in the Introduction, one can conclude from Fig.
\ref{pav} (b) that the peak is related to a solid-hexatic
transition. In this case the BKT transition occurs at
$\rho_{BKT}$=1.508 with unobservable essential singularity at
$\rho_{BKT}$ and a nonuniversal model-dependent  "bump" below
$\rho_{BKT}$ associated with  the entropy liberated by the
unbinding of dislocation pairs. It is interesting to note, that
due to this very thorough averaging we do not observe a possible
$\delta$ function peak related to a first-order hexatic-liquid
transition.

\begin{figure}
\includegraphics[width=8cm]{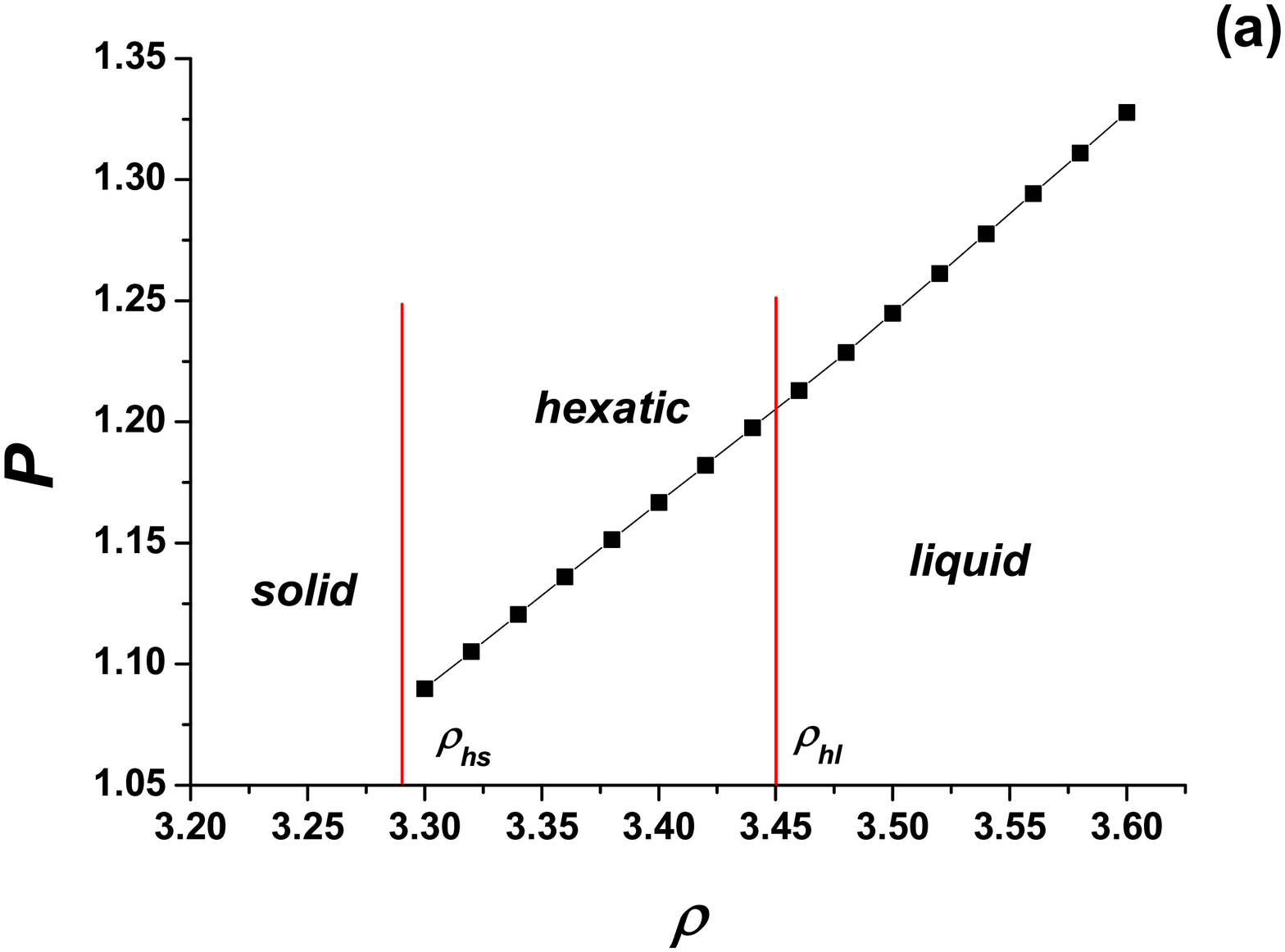}%

\includegraphics[width=8cm]{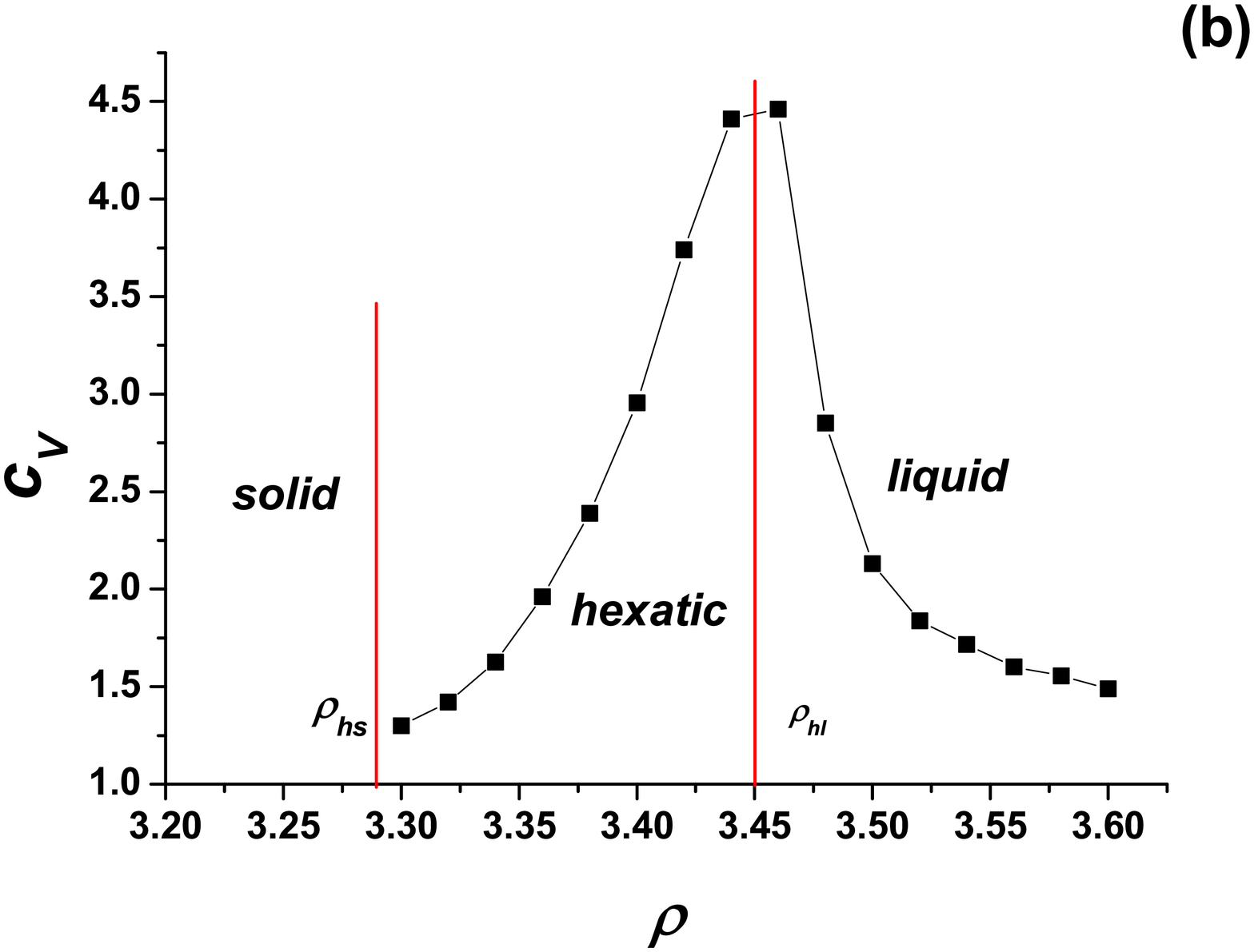}%

\caption{\label{eos-b} (a) The equation of state and (b) heat
capacity of the system in region $b$. $T$=0.003.}
\end{figure}

\begin{figure}
\includegraphics[width=8cm]{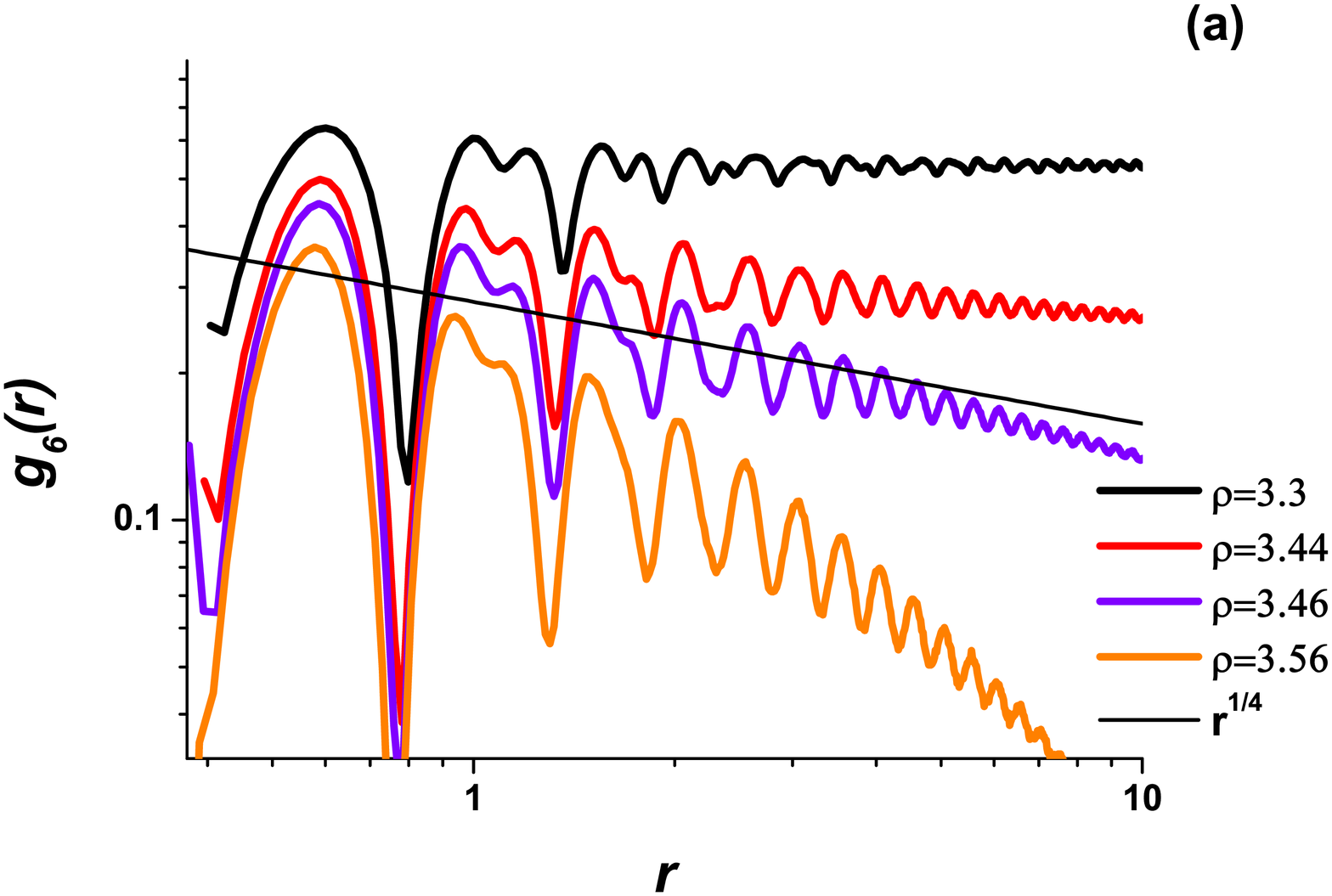}%

\includegraphics[width=8cm]{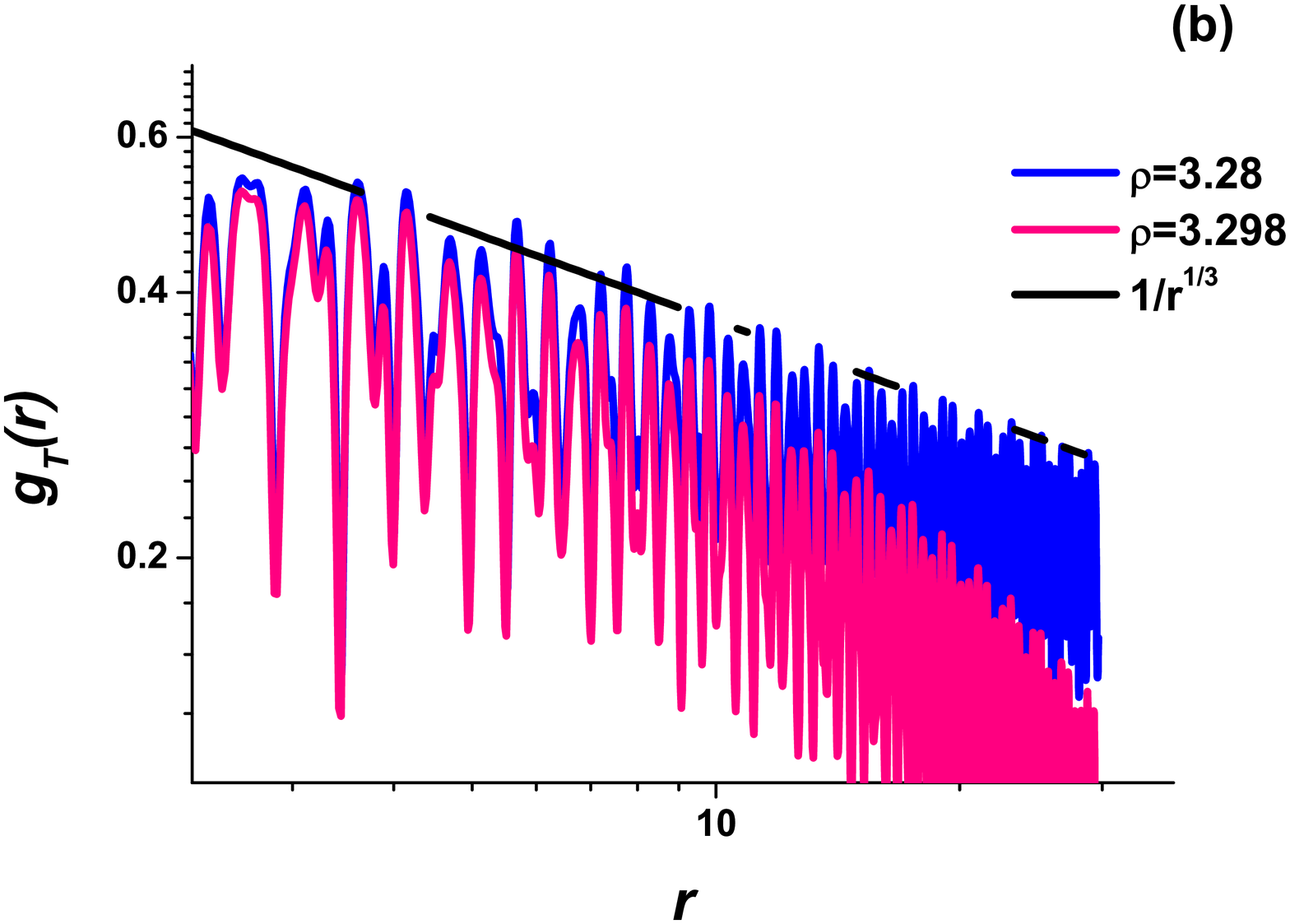}%

\caption{\label{corr-b} (a) The orientational and (b)
translational correlation functions in region $b$. $T$=0.003.}
\end{figure}

We proceeded by a more detailed investigation of region $b$ (see
Fig.~\ref{small}). Here the equation of state and  heat capacity
are less noisy and 50 replicas of the system are sufficient to get
reasonable statistical averaging. Figs.~\ref{eos-b} (a) and (b)
show the equation of state and the heat capacity of the system in
region $b$ at temperature $T$=0.003. One can see that the equation
of state does not demonstrate any loops, therefore we do not
observe a first order phase transition. The system melts in
accordance with the BKTHNY scenario. Figs.~\ref{corr-b} shows the
OCF (panel a) and TCF (panel b) of the system in region $b$. One
can see that the crystal to hexatic transition takes place at
$\rho_{sh}$=3.28 and from the hexatic to liquid phase at
$\rho_{hl}$=3.45. It is also shown that the maximum of heat
capacity is located in the neighborhood of the hexatic phase
stability limit and near the hexatic to liquid transition.
%The peak of the heat capacity takes place
%at the same density $\rho=3.45$.

\begin{figure}
\includegraphics[width=8cm]{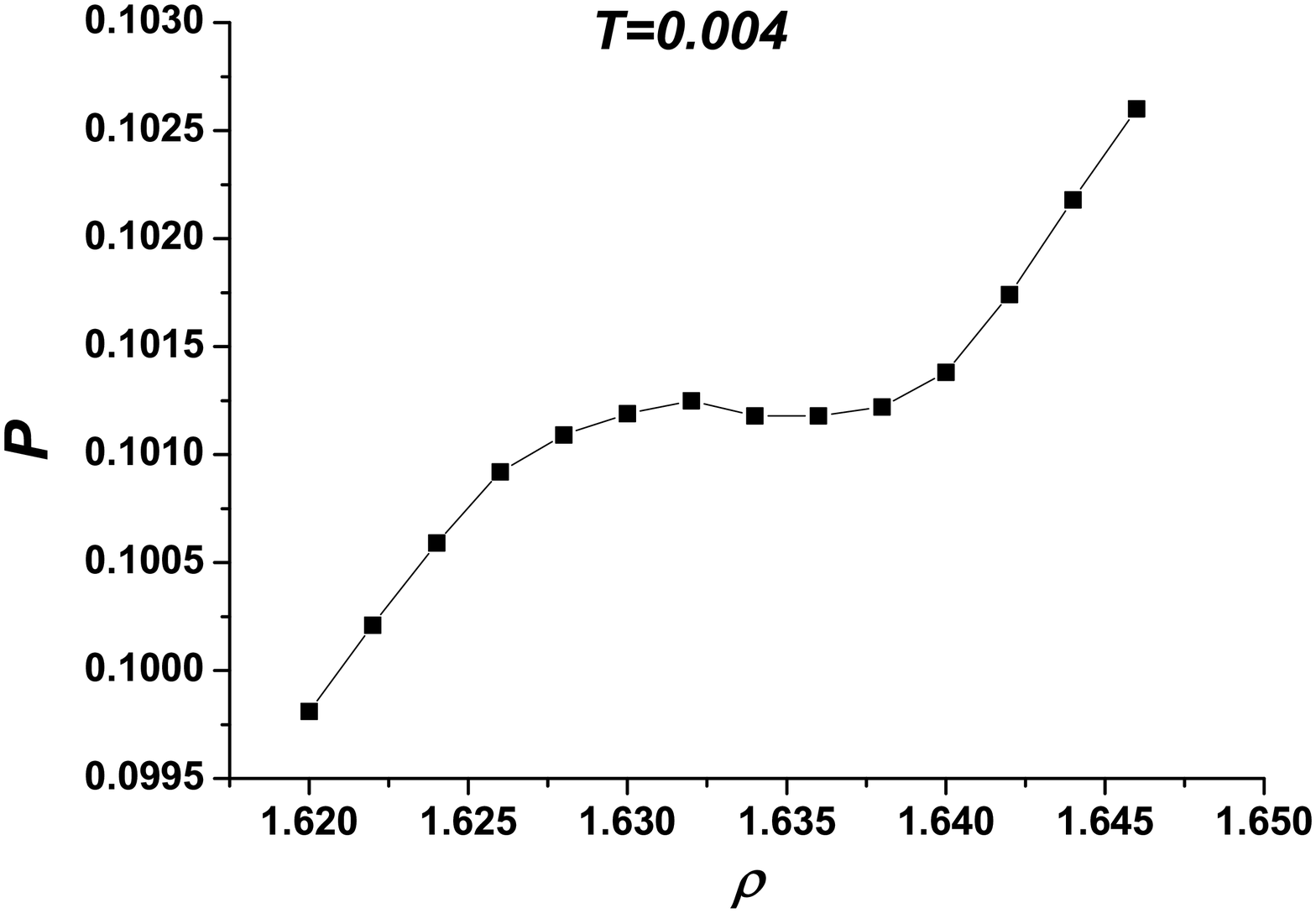}%

\includegraphics[width=8cm]{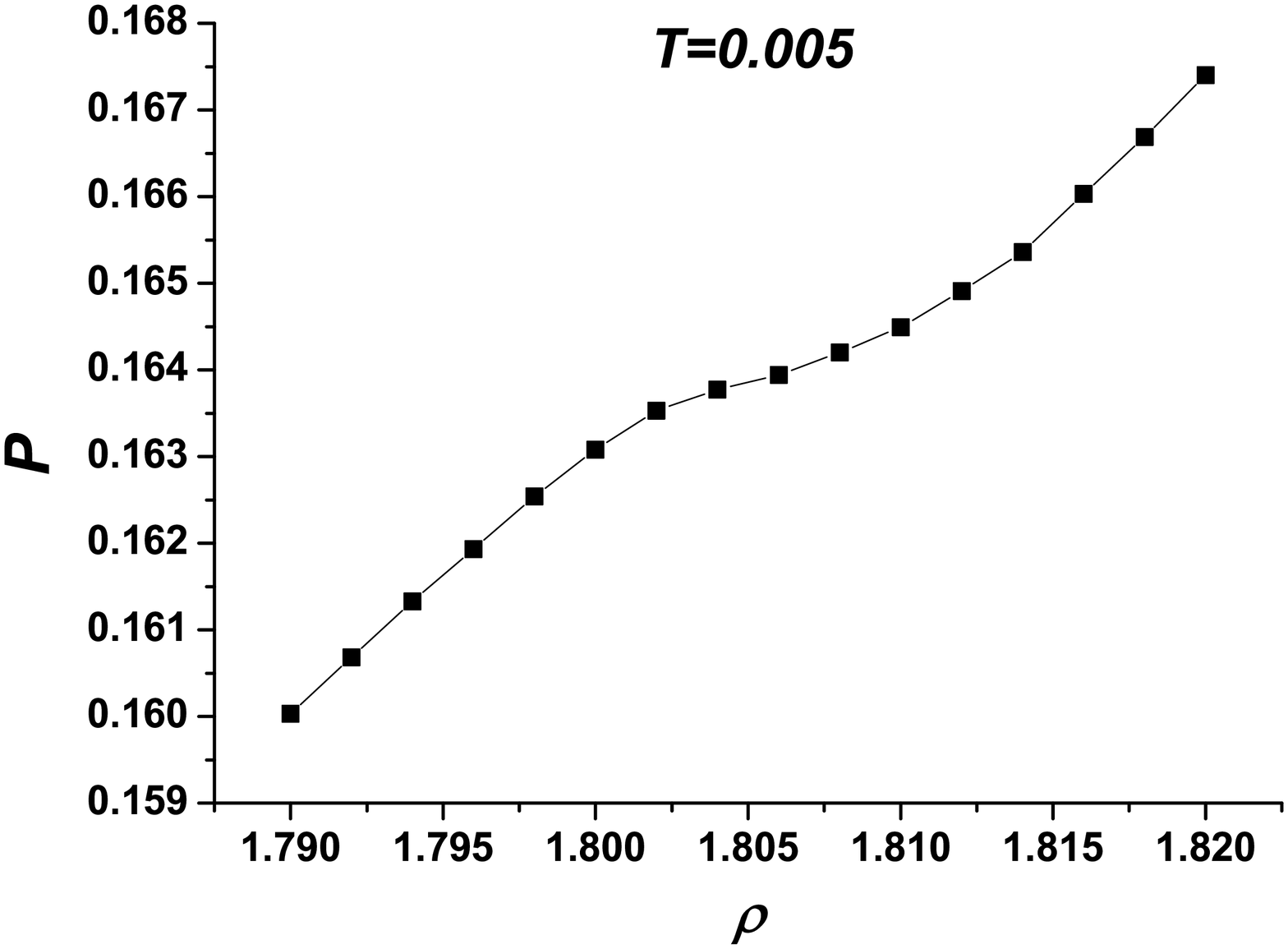}%

\includegraphics[width=8cm]{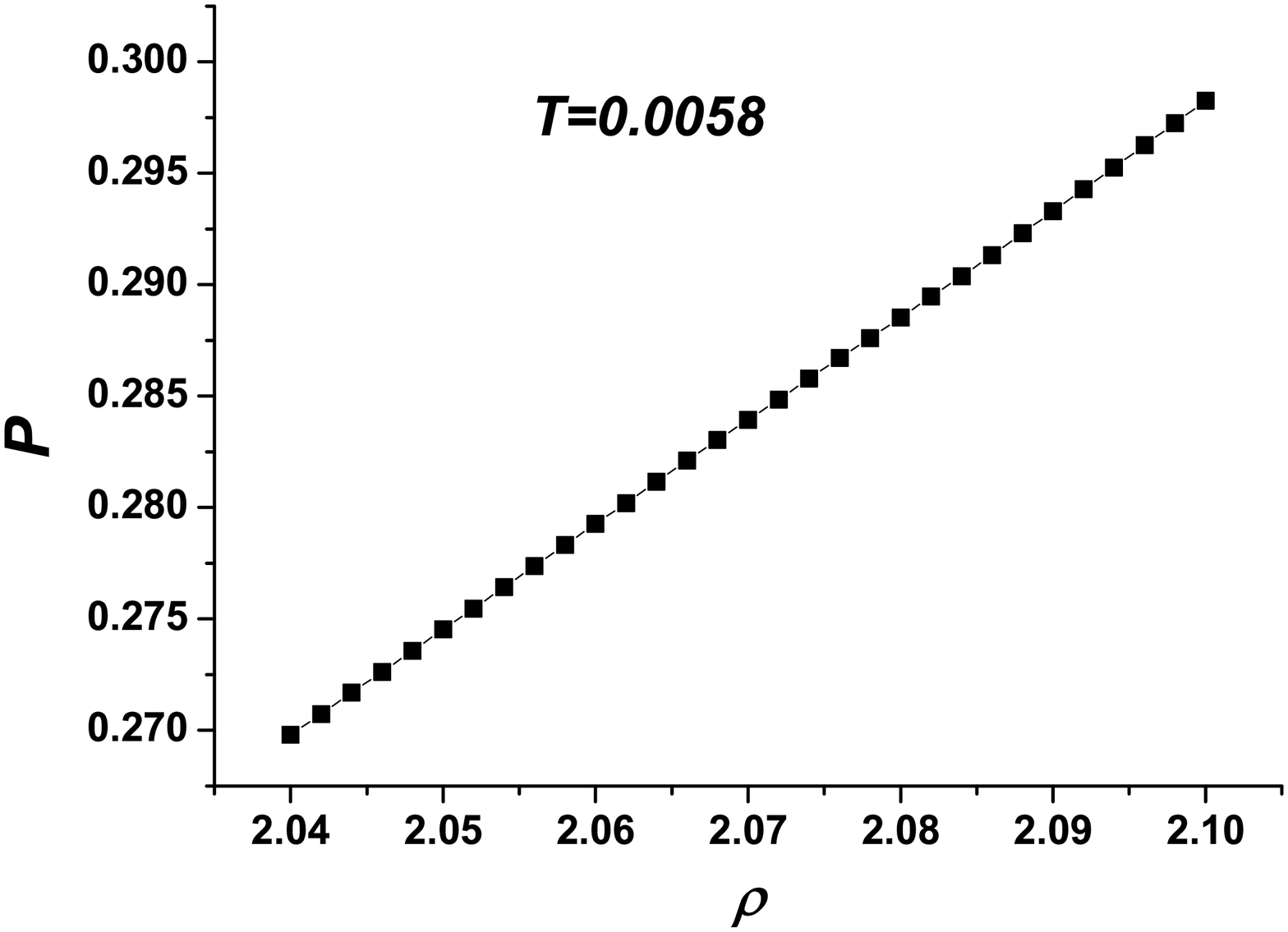}%

\caption{\label{diff-t-a} The isotherms of the system crossing the
melting line in region $a$.}
\end{figure}

Let us consider the melting line in regions $a$ and $b$ at higher
temperatures. Fig. \ref{diff-t-a} shows the equations of state in
region $a$ at three different temperatures. One can see the
Mayer-Wood loop $T$=0.004. However, at $T$=0.005 the Mayer-Wood
loop disappears. Therefore, the melting scenarios change and
melting proceeds in accordance with the BKTHNY theory. No loop is
observed at $T$=0.0058.

A system of Hertz spheres with $\alpha$=5/2 was studied in our
previous work \cite{molphys}. It was shown that the melting line
of the low density triangular crystal had two tricritical points:
at the maximum on the melting line and at $T$=0.0034 at the right
branch (region $b$ in the notation of the present paper). The
results of this paper demonstrate that Hertz spheres with
$\alpha$=7/2 have only one tricritical point at the left branch
(region $a$). No changes of the melting scenarios take place at
the maximum of the melting line.

\begin{figure}
\includegraphics[width=8cm]{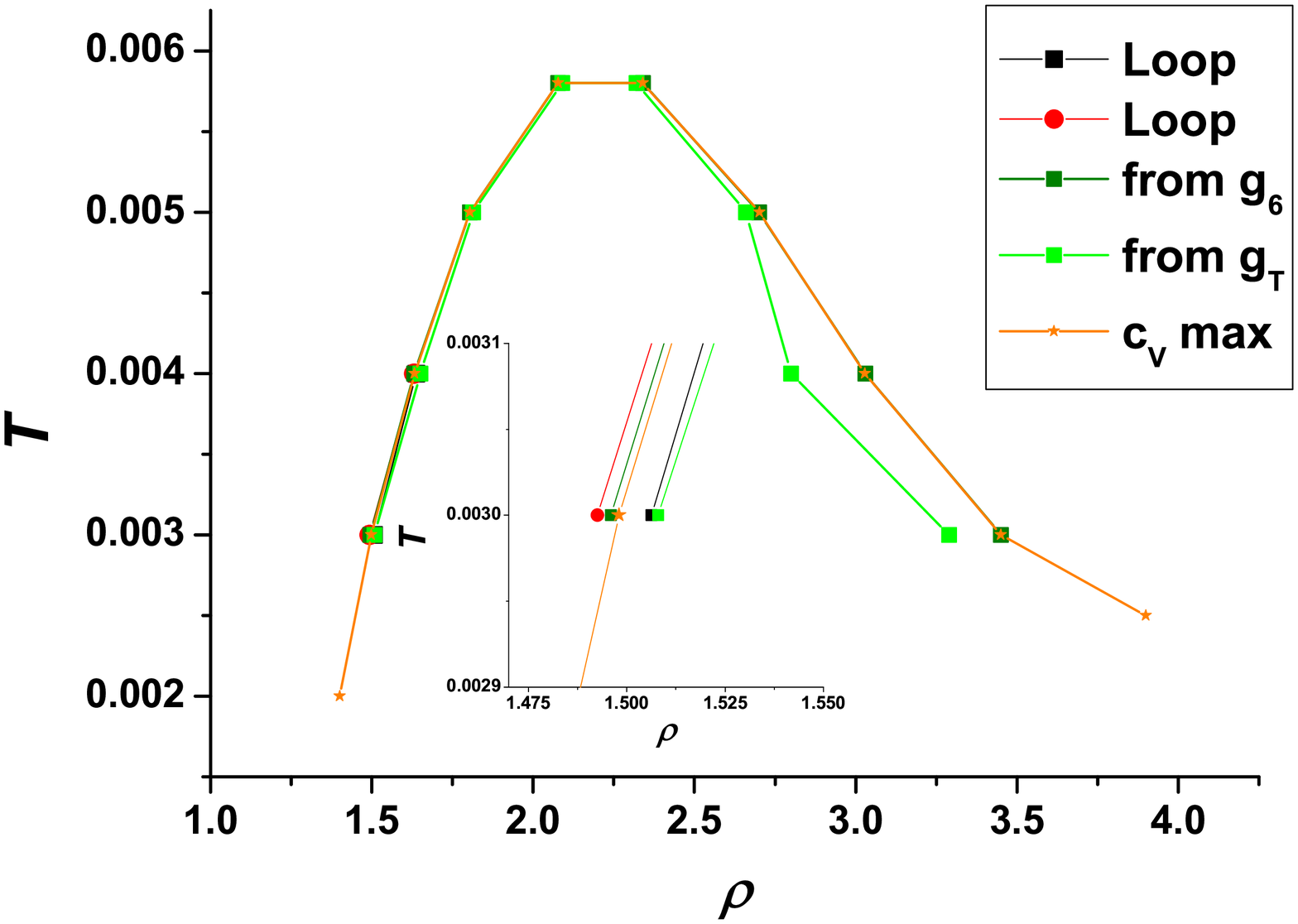}%

\caption{\label{ml-ab} The melting line of a low density
triangular crystal (regions $a$ and $b$). The inset enlarges
 region a at $T$=0.003. The location of the heat capacity maxima is also shown.
 The loop - is the points obtained from the Mayer-Wood loop, $g_6$ - is a solid to hexatic transition from the OCF, $g_T$ -
is a hexatic to isotropic liquid transition from the TCF, $c_V
max$ - is the location of  the heat capacity maxima.}
\end{figure}

The melting line of a low density triangular crystal is shown in
Fig.~\ref{ml-ab}. The location of heat capacity maxima is also
shown. One can see that the heat capacity maximum almost coincides
with the disappearance of orientational order in accordance with
the OCF. In the case of a single first order hexatic to liquid
transition it is located in a two-phase region (see, for example,
Fig.~\ref{pav} (b)). In the case of the BKTHNY scenario a heat
capacity maximum is located near the hexatic to isotropic liquid
transition (Fig.~\ref{eos-b}). But in both cases the heat capacity
maximum is close to the hexatic phase stability limit.
%Because of this it looks that the peak of the heat capacity is related to the hexatic to liquid transition.
It is interesting, that the heat capacity demonstrates peaks even
at $T$=0.0025 and $\rho$=3.9. This temperature is below the
minimum of the melting line and is deep inside the solid phase.
%, and therefore no liquid phase exists.
The location of the heat capacity peak seems to coincide with the
density of the minimum of the melting line. Also the height of the
maximum is much lower compared with the peaks related to the
melting transition. One can suppose that in the vicinity of the
melting line minimum there is smooth structural reconstruction
producing this maximum.
%conclude that approaching to the transition to liquid phase induces the increase of the specific heat.

%%%%%%%%%%%%%%%%%%%%%%%         cccccccccccc

\begin{figure}
\includegraphics[width=8cm]{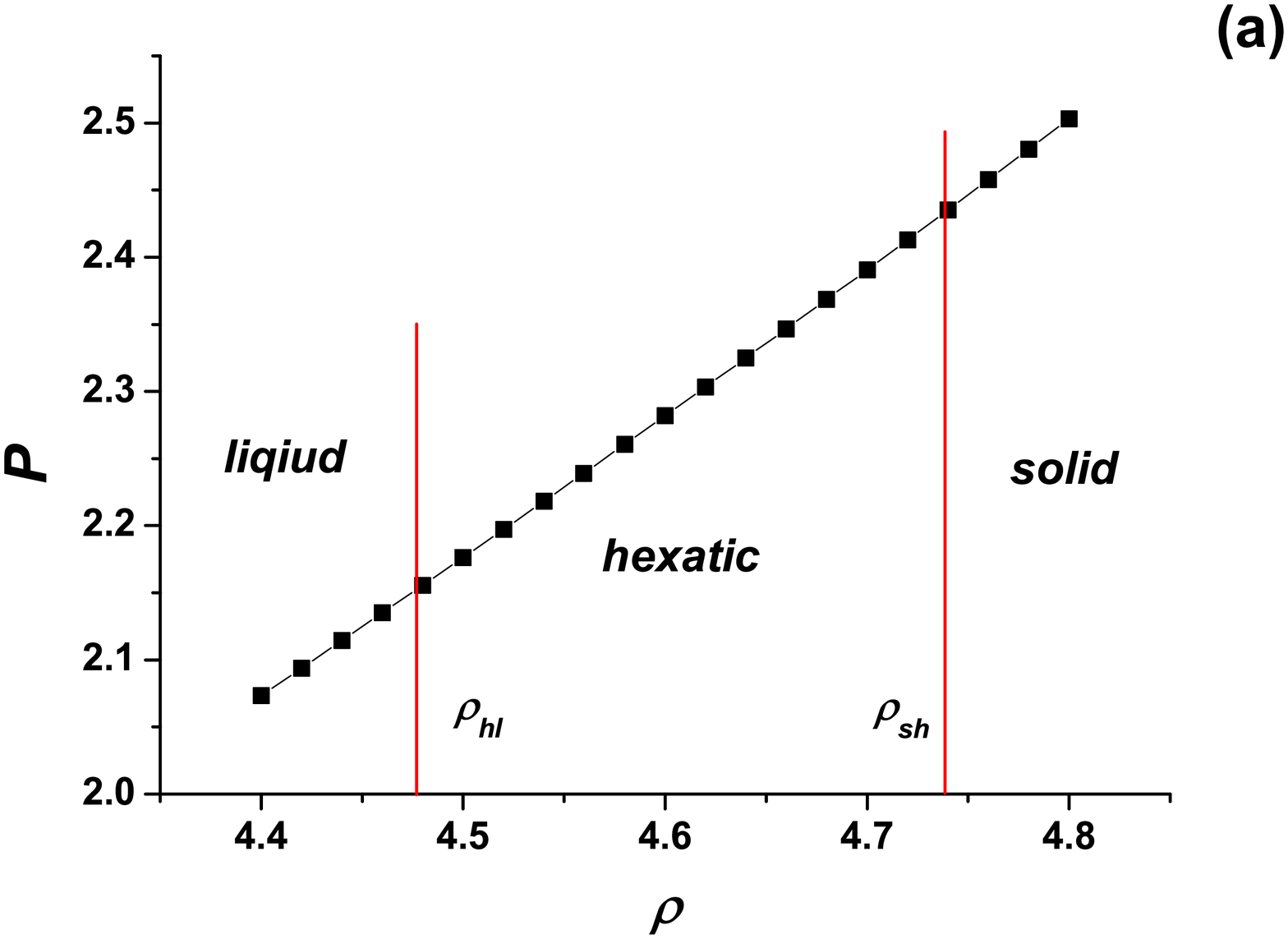}%

\includegraphics[width=8cm]{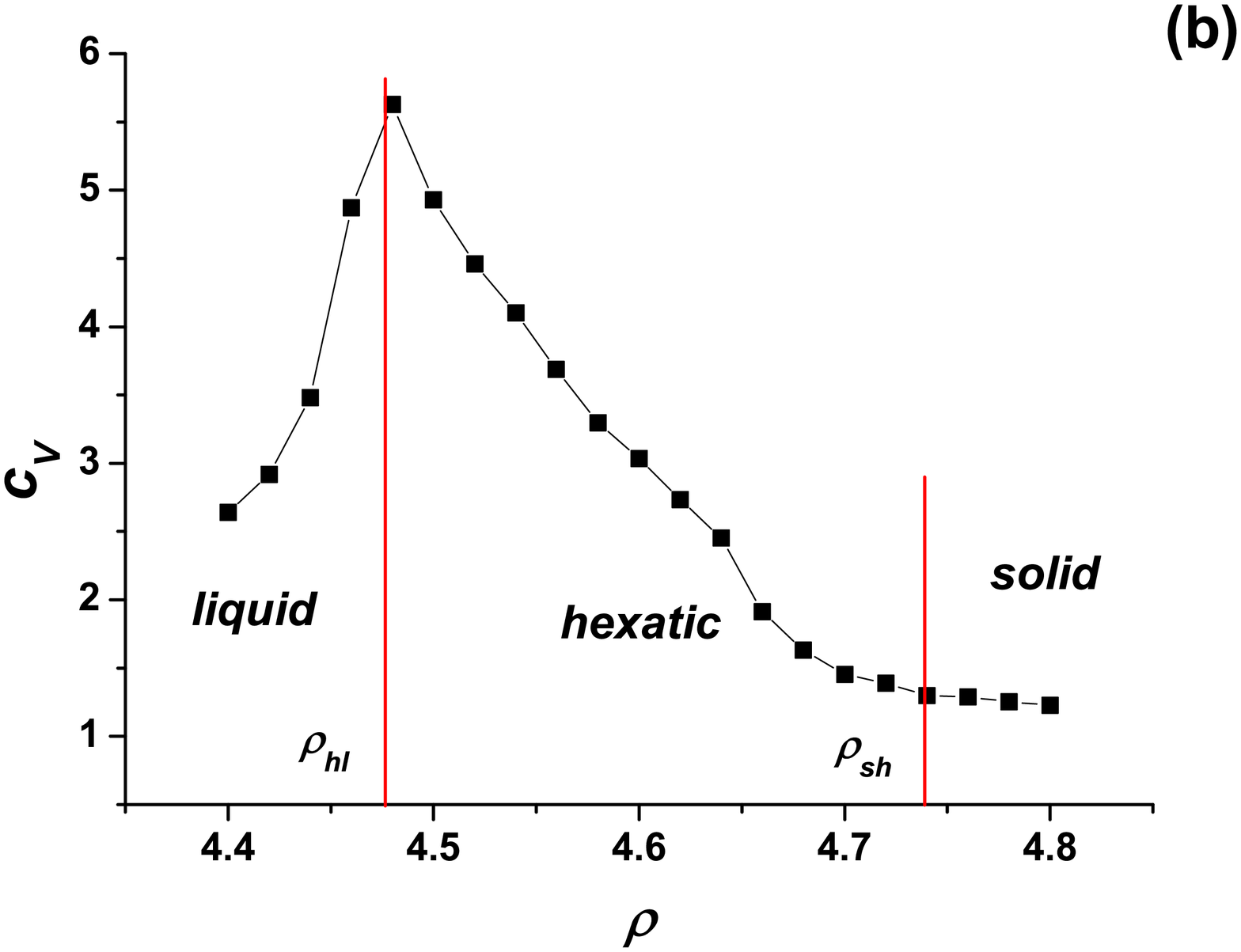}%

\caption{\label{eos-c} (a) The equation of state and (b) heat
capacity of the system in region $c$ at $T$=0.003.}
\end{figure}

We proceed with considering region $c$. Here we again performed
simulation of 50 replicas in order to get a reliable statistic.
The equation of state again does not demonstrate the Mayer-Wood
loop (Fig. \ref{eos-c} (a)), therefore the BKTHNY scenario is
realized in this case. The maximum of heat capacity appears at
$\rho$=4.48 (Fig.~\ref{eos-c} (b)).

\begin{figure}
\includegraphics[width=8cm]{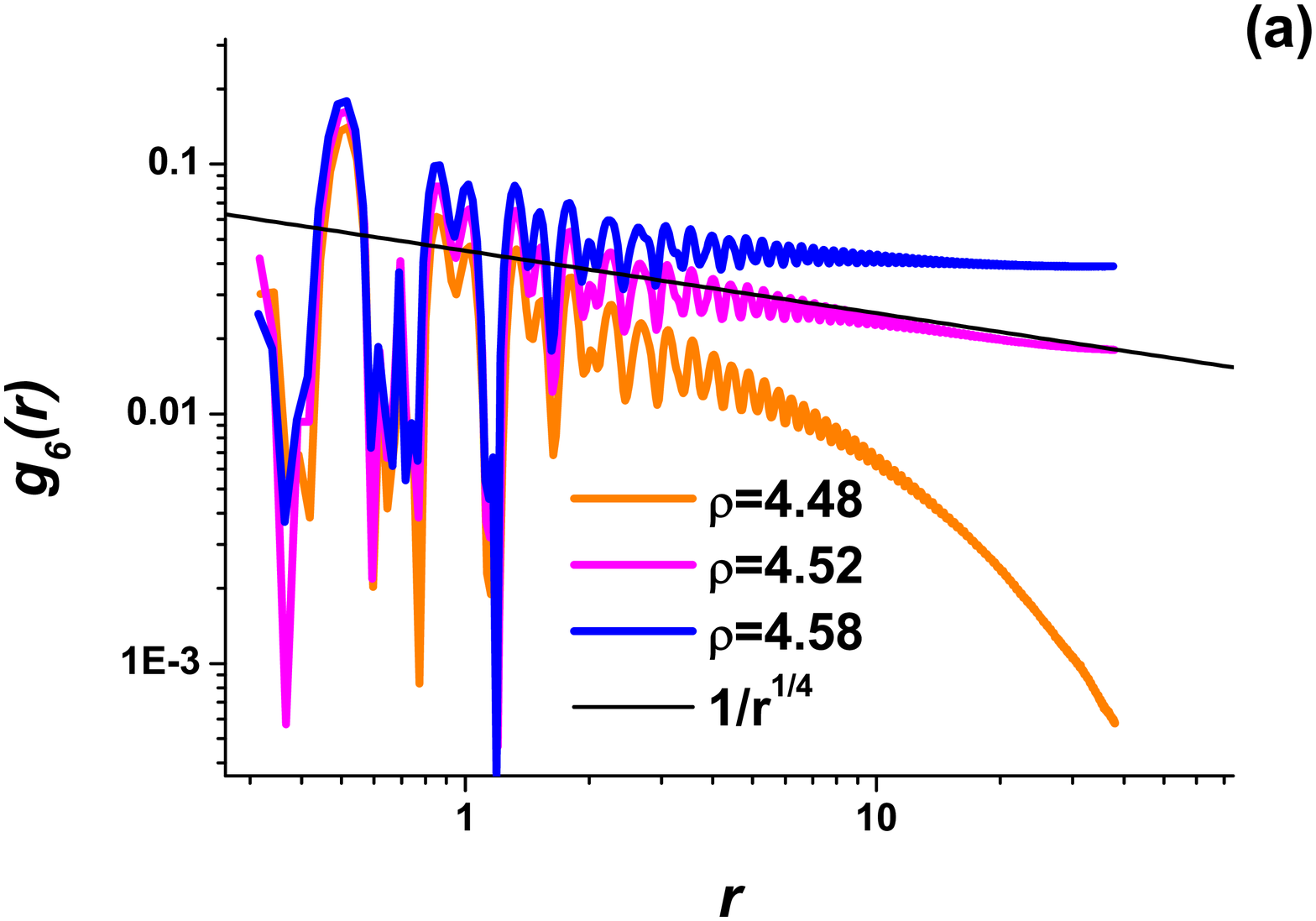}%

\includegraphics[width=8cm]{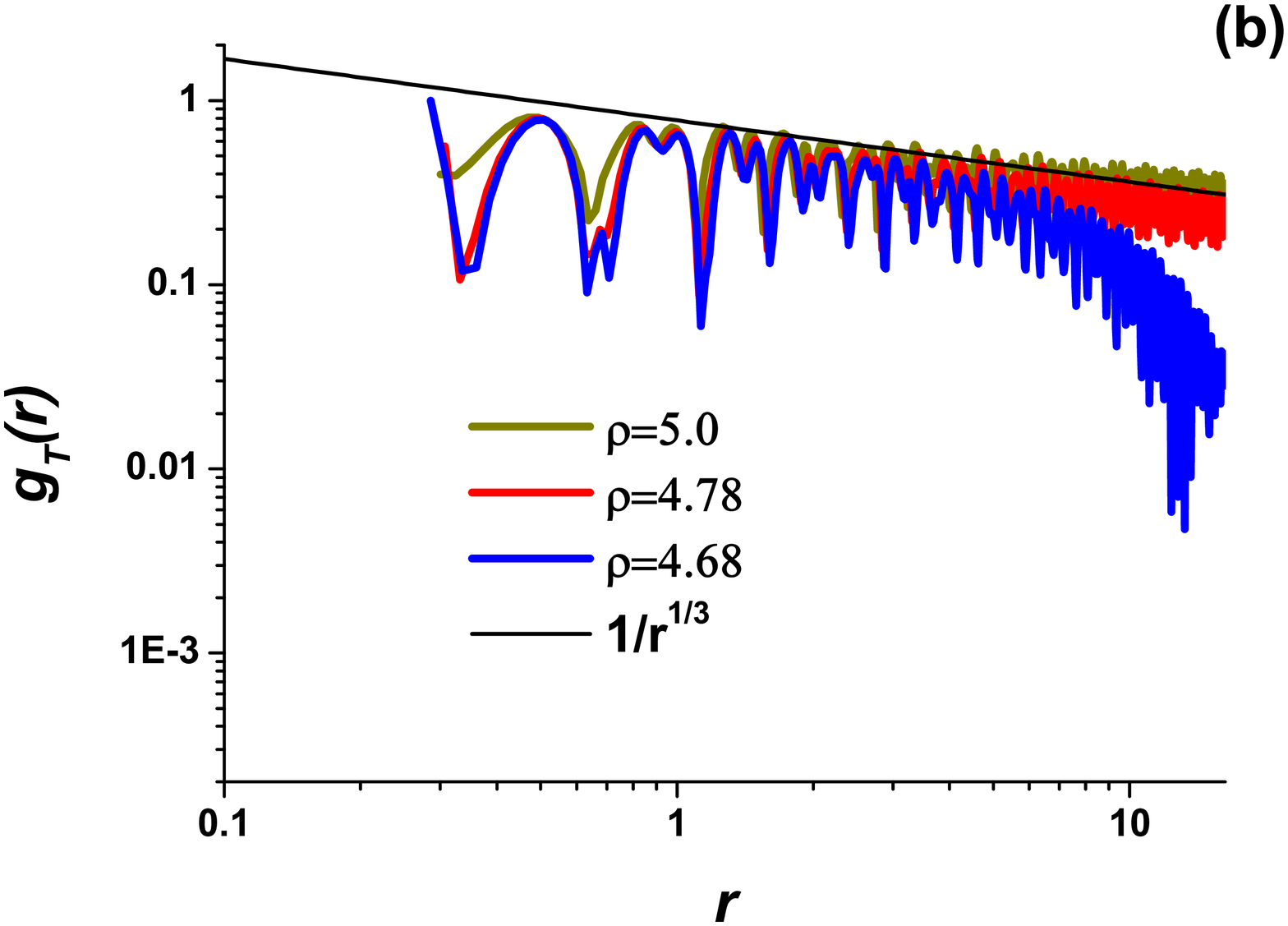}%

\caption{\label{corr-c} ((a) The orientational and (b)
translational correlation functions in region $c$ at $T$=0.003.}
\end{figure}

The limits of stability of the solid and hexatic phases are
obtained from correlation functions $g_6$ and $g_T$
(Fig.~\ref{corr-c}). The solid to hexatic transition takes place
at density $\rho_{sh}$=4.74 and the hexatic to liquid one at
$\rho_{hl}$=4.472. It means that the peak of heat capacity appears
close to the hexatic to liquid transition.

Finally we studied the system in region $d$. The equation of state
at $T$=0.003 does not demonstrate the Meyer-Wood loop, and one can
conclude that in this region the system melts in accordance with
the BKTHNY scenario. The situation is similar to regions $b$ and
$c$. From the long-range behavior of the OCF and TCF one can see
that the transition from solid to hexatic phase takes place at
$\rho_{sh}$=7.36 and the transformation of the hexatic phase into
the isotropic liquid occurs at $\rho_{hl}$=7.42.

In Ref. \cite{miller} it was found that the melting line of a
Hertzian system with $\alpha$=7/2 had a complex shape with
multiple maxima and minima. Our calculations confirm this unusual
result. However, the nature of such a complex shape of the melting
line of the same phase remains vague. To shed light on this
problem we studied the equations of state and the structure of the
solid phase in the region of densities from b to c below the
melting line.

\begin{figure}
\includegraphics[width=8cm]{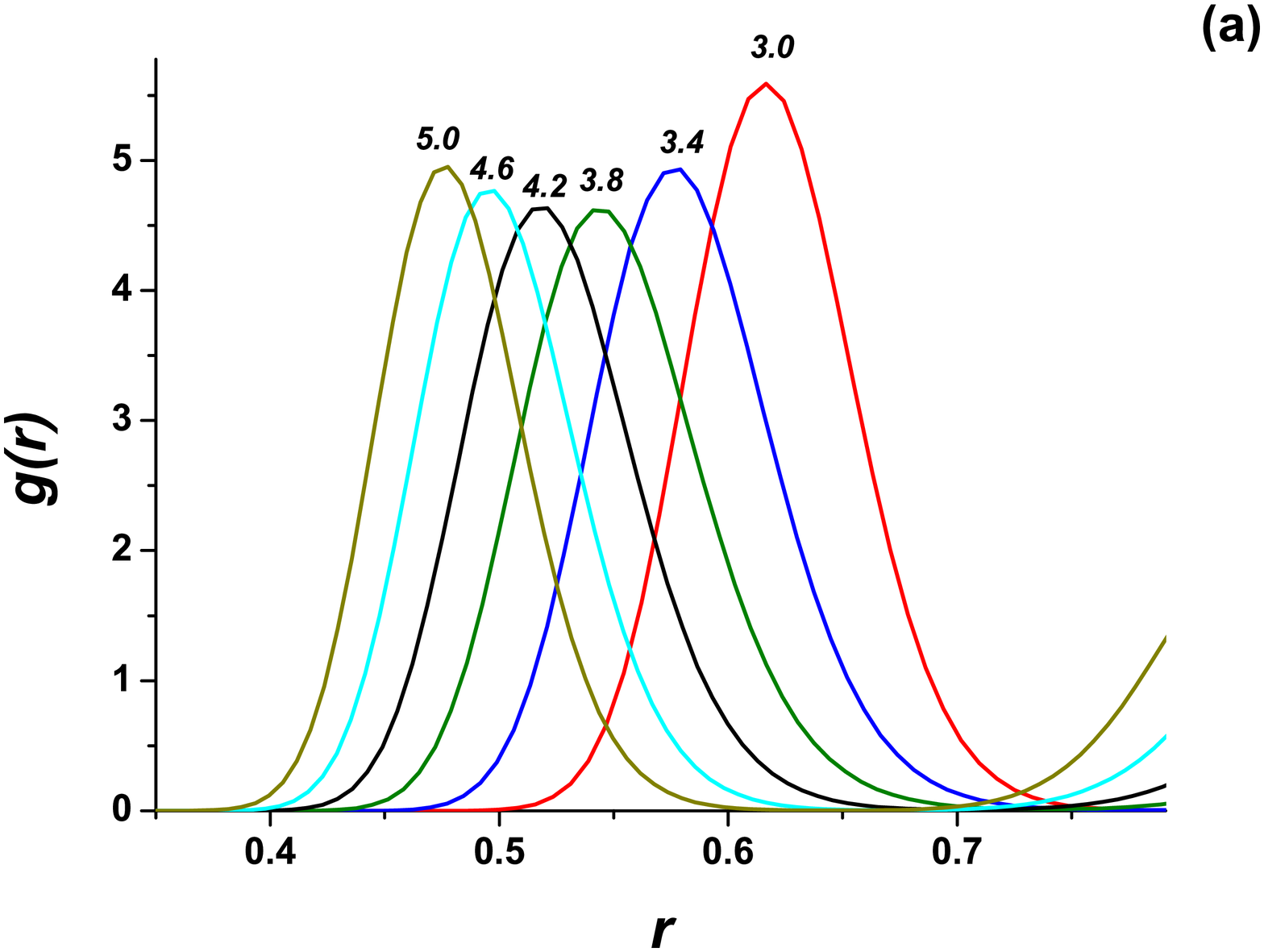}%

\includegraphics[width=8cm]{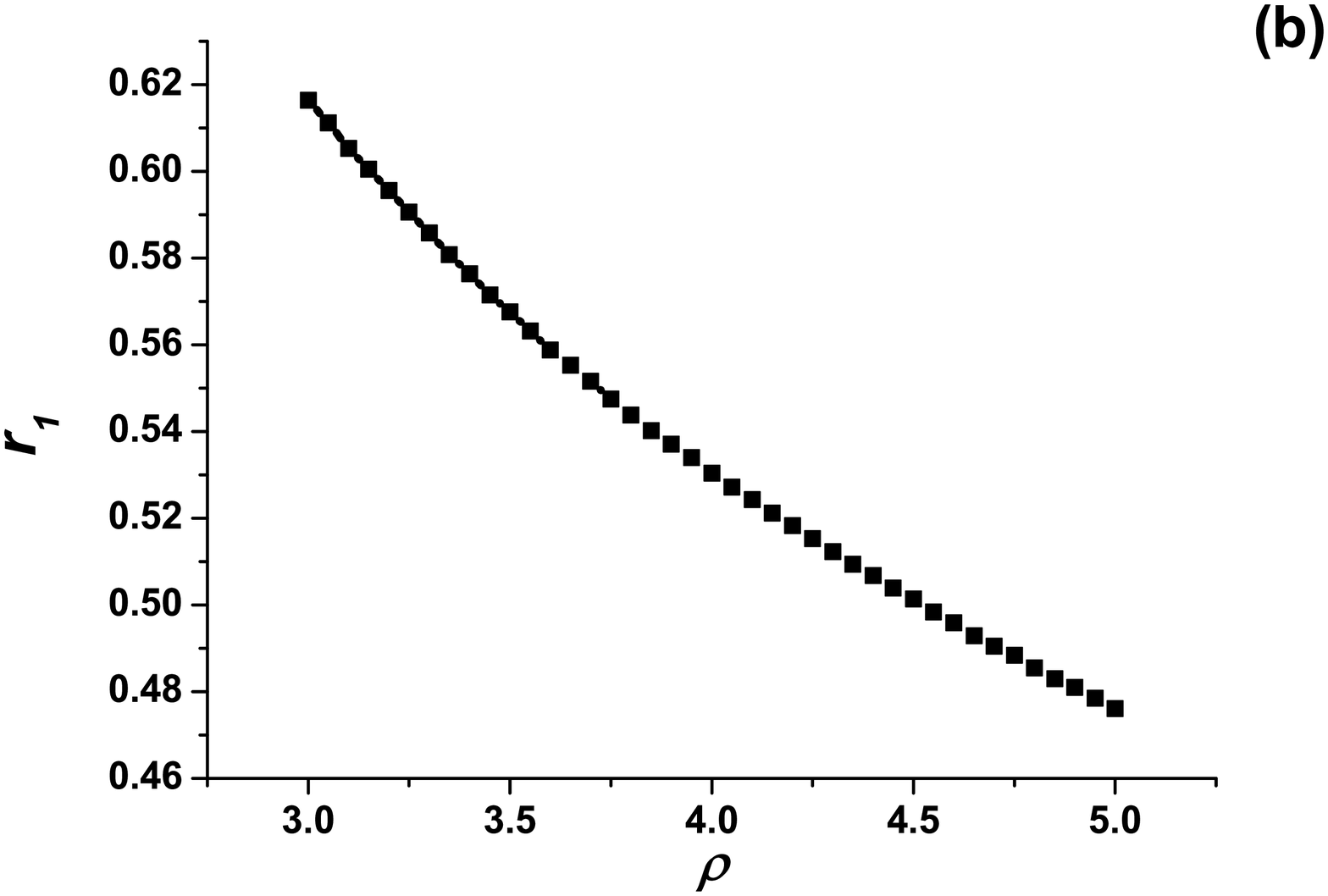}%

\includegraphics[width=8cm]{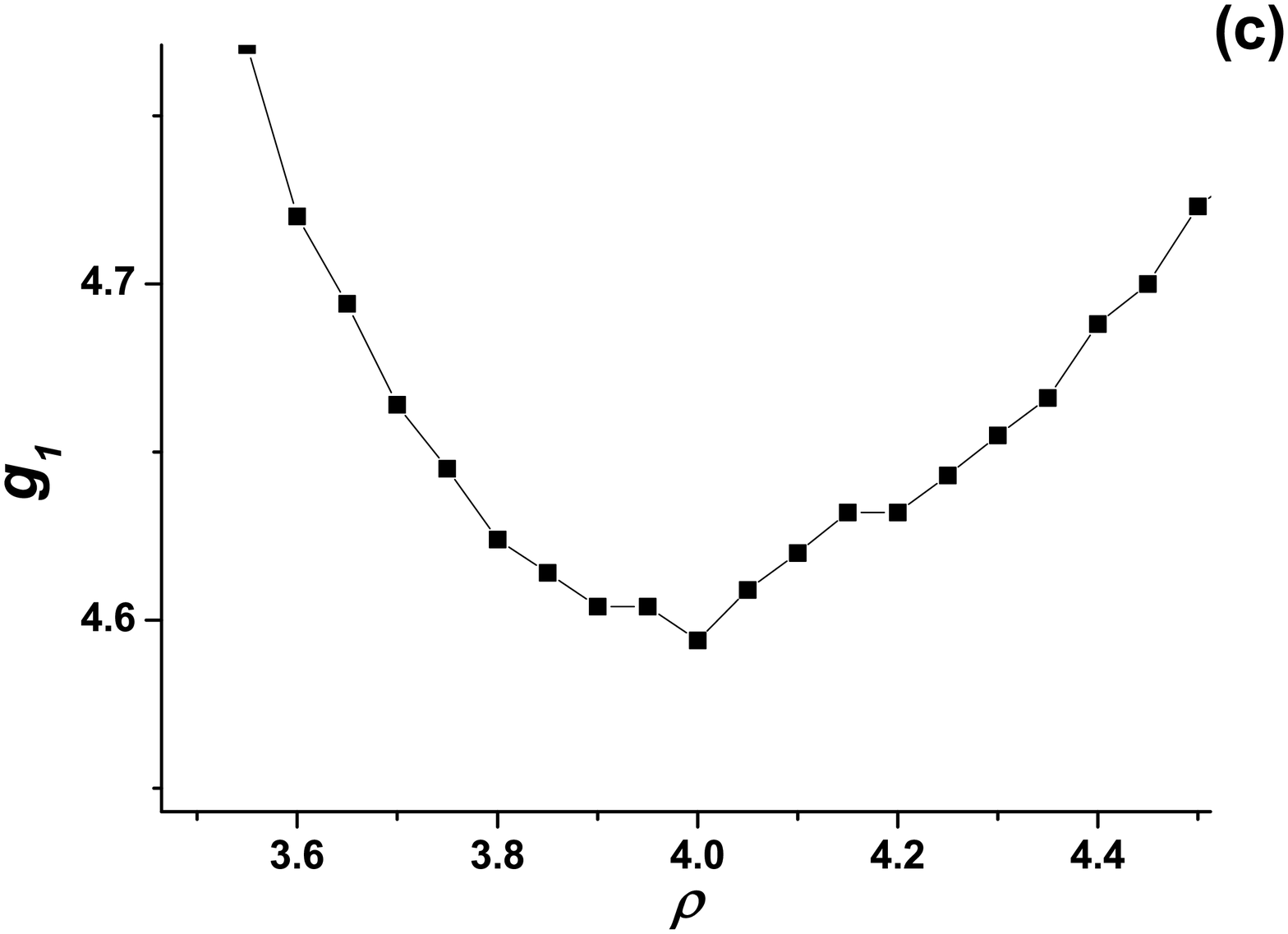}%

\caption{\label{rdf-t002} (a) The first peak of the rdf at
$T$=0.002. (b) The location of the first peak of the rdf at
$T$=0.002. (c) The height of the first peak of the rdf at
$T$=0.002.}
\end{figure}

Fig.~\ref{rdf-t002} (a) shows the first peak of the rdf at
$T$=0.002 as a function of density. Usually with an increase in
density the location of the first peak shifts towards smaller
$r$'s while its height increases. Here we clearly see that
although the location of the first peak behaves normally (Fig.
\ref{rdf-t002} (b)) its height decreases with density increasing
to $\rho$=4.0 where a minimum is observed (Fig. \ref{rdf-t002}
(c)). This density of the minimum is close to the density of the
melting line minimum. Such unusual behavior of the rdfs is well
known in liquids where it signalizes smooth structural crossover,
i.e. the structure changes without a phase transition. A
structural crossover in liquids usually leads to the appearance of
a maximum on the melting line (see, for example, \cite{book1}). In
the present work the situation is more complex: there is no change
in the structure (only a triangular crystal is observed), however,
the unusual dependence of its rdfs leads to the appearance of a
minimum on the melting line.

\begin{figure}
\includegraphics[width=8cm]{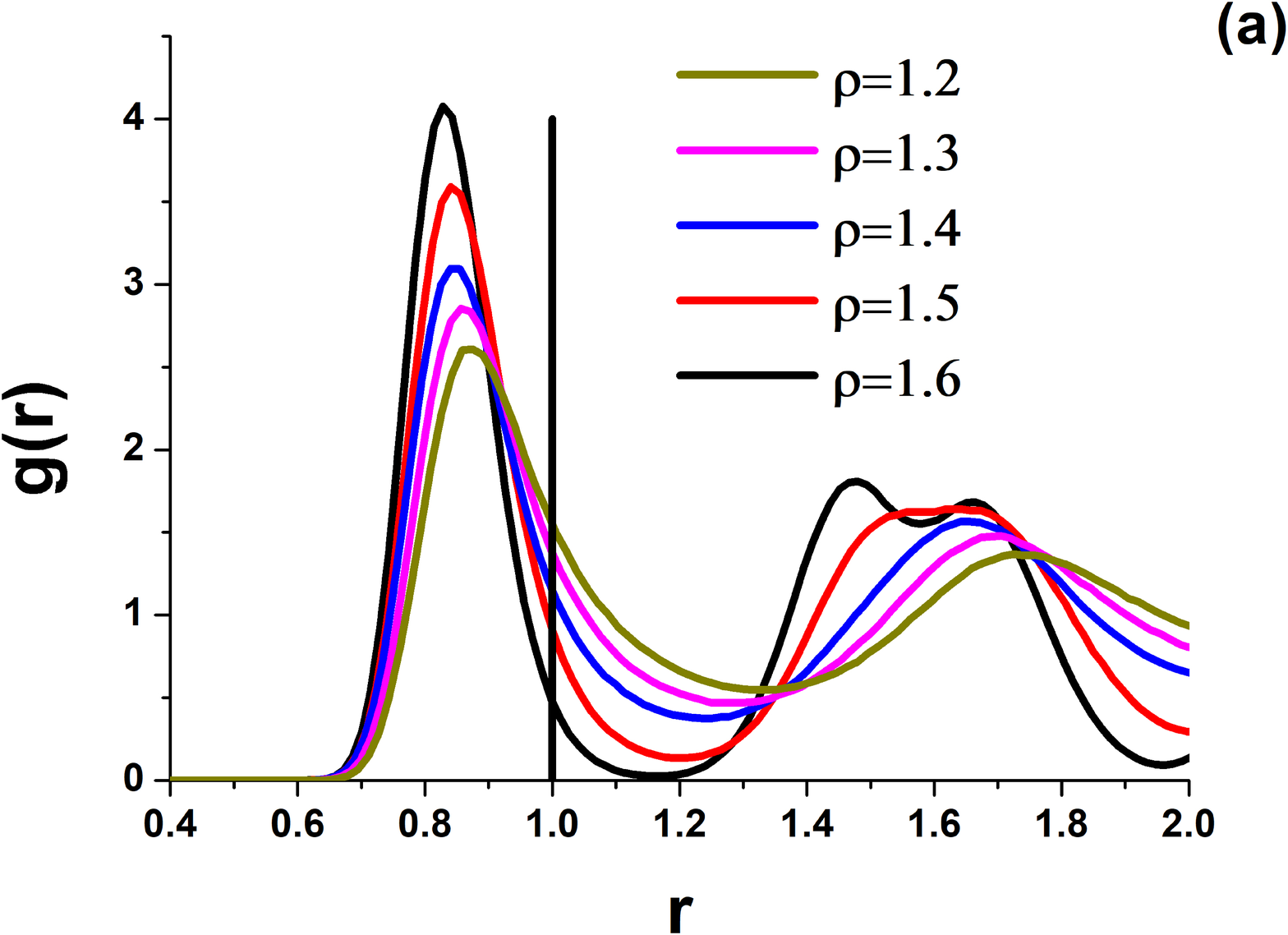}%

\includegraphics[width=8cm]{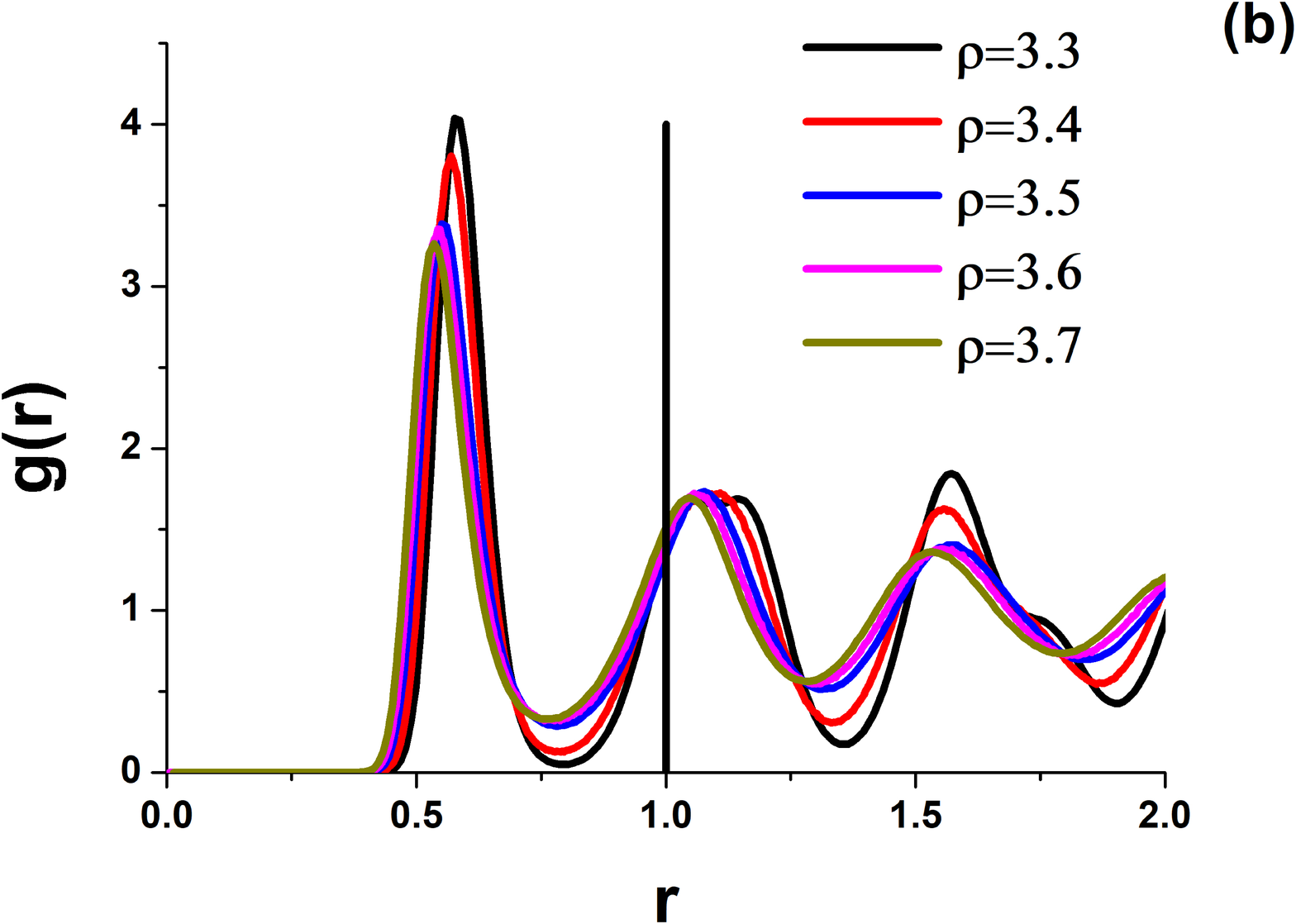}%

\includegraphics[width=8cm]{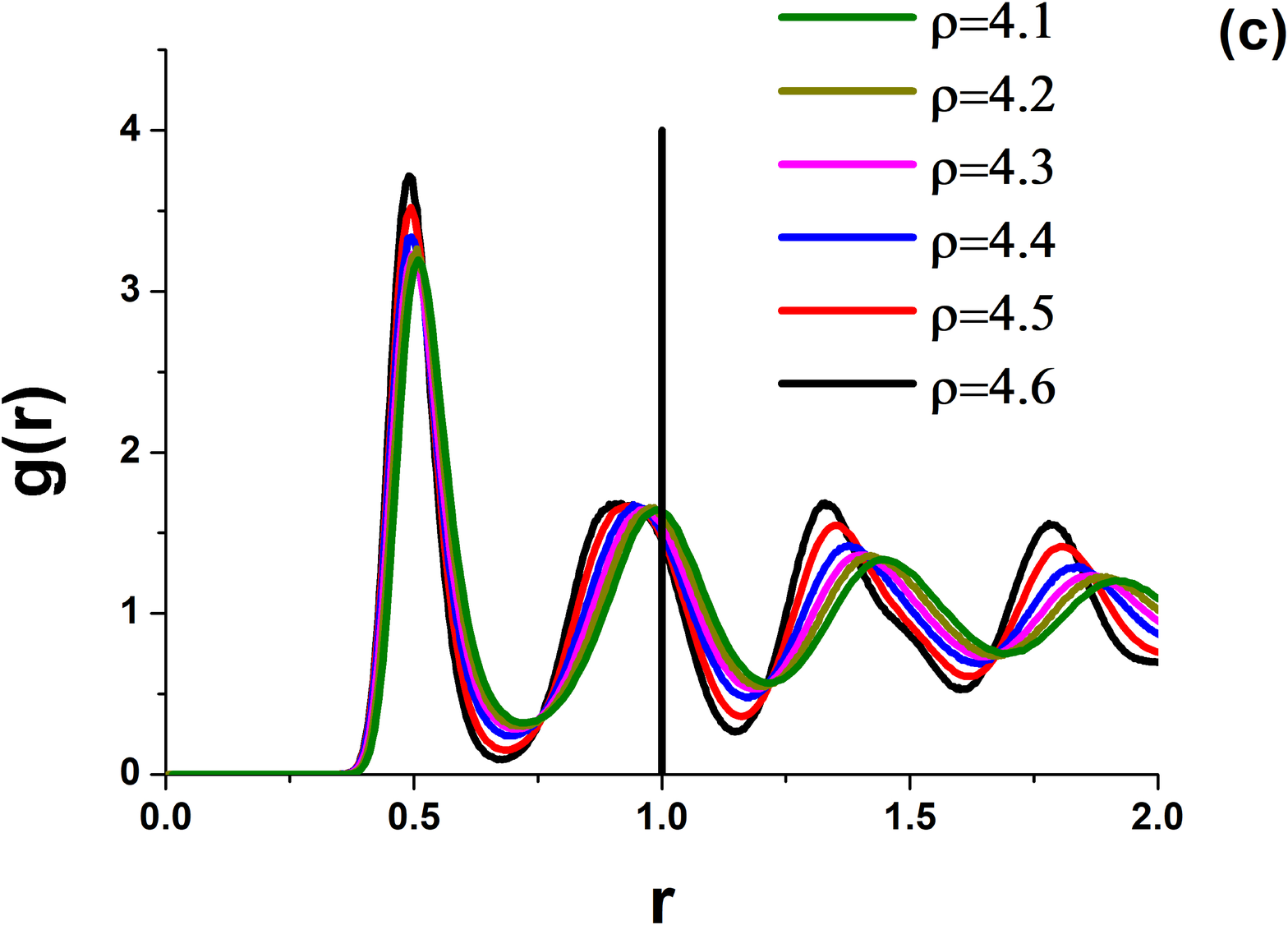}%

\includegraphics[width=8cm]{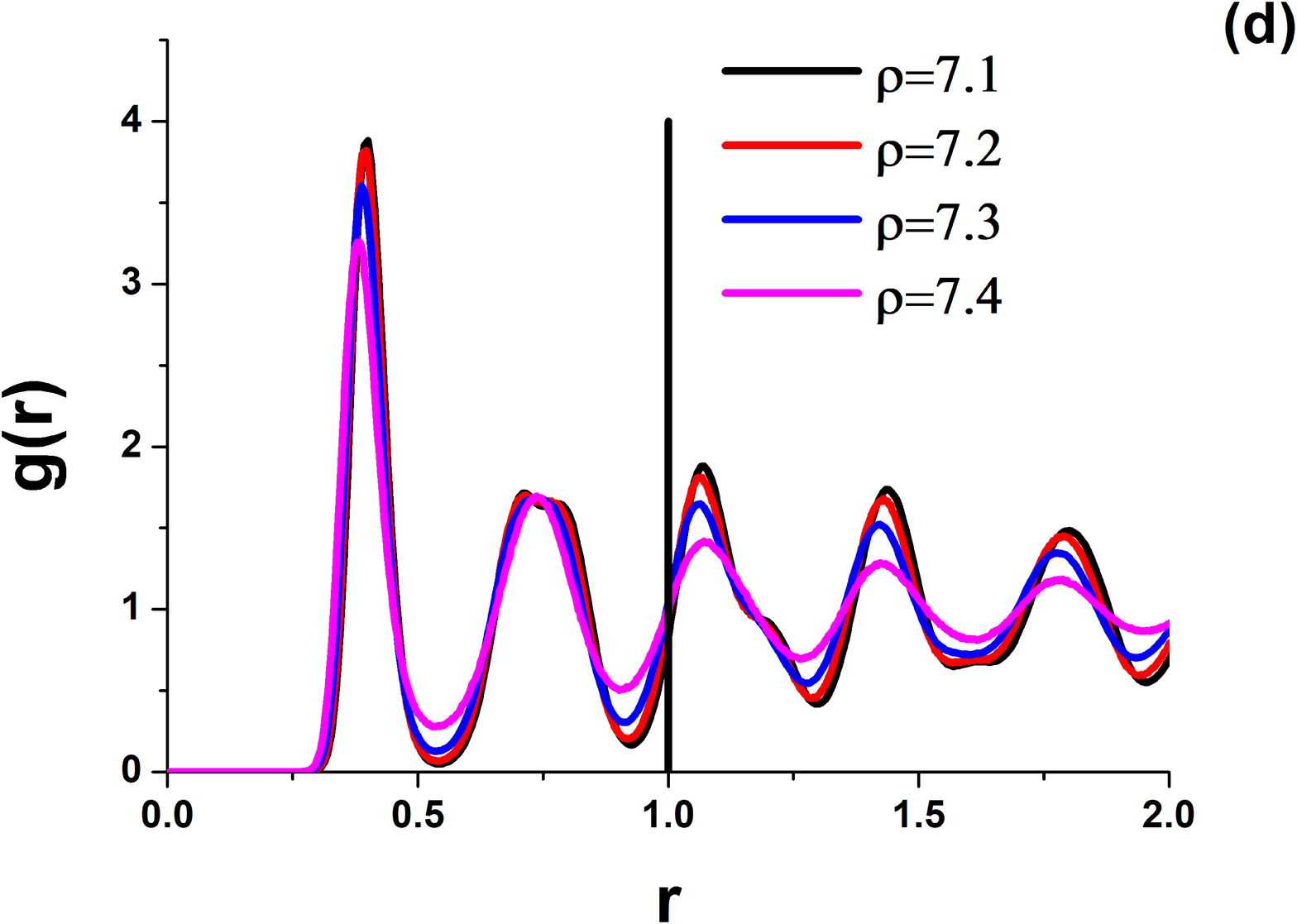}%

\caption{\label{rdf-abcd} The radial distribution functions along
the $T$=0.003 isotherms in the vicinity of the regions of phase
transitions. Panels a, b, c and d correspond to regions $a$, $b$,
$c$, and $d$. The vertical line shows the $r$=1.0 threshold, which
is the realm of influence of the potential.}
\end{figure}

Another unusual property of the phase diagram is its
quasi-periodic shape. In the density interval up to $rho$=10.0
which was studied in \cite{miller} there are three peaks. Such
complex phase diagrams can be observed in core-softened systems
with several length scales in the potential
\cite{dfrt1,dfrt2,dfrt3,dfrt5,we5}. In the case of deformable
Hertzian potential there is only one length scale. Here we
suppose, that complex phase behavior can be related to the absence
of singularity of the potential in the origin and to high
compressibility of the system. As a result of these two factors,
more coordination spheres come into the realm of potential
influence when the system is compressed. Fig.~\ref{rdf-abcd} shows
the rdfs of the system in the vicinity of the phase transition
regions. Regions $a$ and $c$ correspond to normal melting, i.e.
the melting curve has a positive slope in these regions (a solid
is denser than a liquid), while regions $b$ and $d$ are re-entrant
melting regions, i.e. the liquid has higher density than the
solid. The vertical line shows the potential cut-off distance. One
can see that in region $a$ the first peak of the rdf is within the
potential range while the first minimum is out of it. In the
vicinity of region $b$ the first minimum of the rdf is within the
potential cut-off distance, while the second maximum is still out
of potential cut-off. The system crystallizes again in region $c$
where the second peak is within the interaction range. Finally, in
region $d$ the second minimum is inside the potential cut-off
distance. From these observations we see that when a peak of the
rdf occurs in the distance within the interaction range ($r$=1.0)
the system crystallizes, while when a minimum of the rdf takes
place in this realm the system experiences re-entrant melting.
Therefore, crystallization of the system is strongly influenced by
the number of coordination shells within the interaction range.

In order to evaluate the influence of different coordination
shells we performed the following procedure. In the case of an
ideal triangular lattice the relation between the density and the
lattice parameter is $a=\rho \sqrt 3/2$. The second neighbor
distance is $a \sqrt{3}$ and the third one is $2a$. The particles
interact if the distance between them is below cut off $r_c$=1.0.
Therefore, we can calculate how many coordination shells are
within the interaction range for some density of a triangular
lattice.

\begin{figure}
\includegraphics[width=8cm]{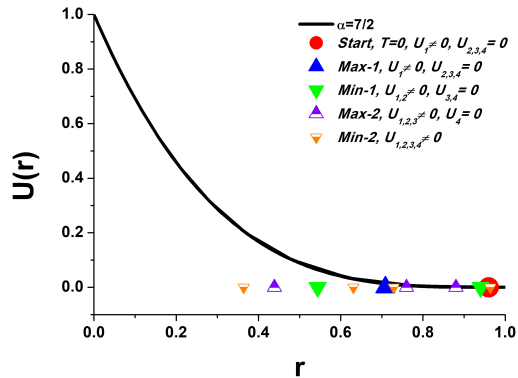}%

\caption{\label{pot-max-min} The interaction potential of a
Hertzian system with $\alpha$ =7/2 with special points (see the
text).}
\end{figure}

At $T$=0 the particles start to interact (i.e. the distance
between the nearest neighbors becomes equal to unity) at density
$\rho$=1.1547 (the Start point in Fig. \ref{pot-max-min}). This
point corresponds to crystallization at branch $a$ of the melting
line. The contribution of the first neighbors to the energy
becomes positive $U_1 \neq 0$, while the contribution of the
highest order neighbors is zero $U_2=U_3=U_4$=0. The $Max-1$ point
corresponds to the first maximum of the melting line (the merging
of the $a$ and $b$ regions). At this point the energy is still
determined by the first coordination shell only. The $Min-1$ point
is at $\rho$=3.9 where the minimum of the melting line takes
place. At this point the second coordination shell comes into the
interaction distance. The second maximum of the melting line is
denoted as $Max-2$. At this point three coordination shells are
within the interaction range. Finally, at the second minimum of
the melting line (point $Min-2$) four coordination shells are
within the cut-off distance. One can see that when an odd number
of coordination shells is within the cut-off distance the melting
line has a positive slope and the solid density is higher than the
density of the corresponding liquid, whilst when the number of
coordination shells becomes even the melting curve passes the
maximum and turns to a negative slope with the liquid density
higher than the solid one. Therefore, the complex shape of the
phase diagram should be related to the inclusion of higher order
neighbors in the interaction range.

\section{Conclusions}

The paper presents a molecular dynamics simulation study of the
phase diagram and melting scenarios of a two-dimensional system of
Hertzian spheres with control parameter $\alpha$=7/2 previously
studied in Ref. \cite{miller}. It is shown that depending on the
position on the phase diagram, the system can melt both in
accordance with the
Berezinskii-Kosterlitz-Thouless-Halperin-Nelson-Young (BKTHNY)
scenario (two continuous transitions with an intermediate hexatic
phase) as well as through a two-stage melting with a first-order
hexatic-isotropic liquid transition and a continuous solid-hexatic
BKT transition. The behavior of heat capacity was studied. We show
that despite two-stage melting, the heat capacity has one peak
which seems to correspond to a solid-hexatic transition. This peak
is a nonuniversal model-dependent maximum below the solid-hexatic
transition density associated with the entropy liberated by the
unbinding of bound dislocation pairs \cite{str,ch_l}. In the case
of the first-order hexatic-liquid transition the heat capacity
peak is located inside the two-phase region while in the case of
the BKTHNY scenario the peak is inside the hexatic phase in the
vicinity of the hexatic-liquid transition. The form of the phase
diagram is related to the number of coordination shells inside the
potential range.

\section{Acknowledgments}
We are grateful to V.V. Brazhkin, and E.E. Tareyeva for
stimulating discussions. This work was carried out using computing
resources of the federal collective usage centre "Complex for
simulation and data processing for mega-science facilities" at NRC
"Kurchatov Institute", http://ckp.nrcki.ru, and supercomputers at
Joint Supercomputer Center of the Russian Academy of Sciences
(JSCC RAS). The work was supported by the Russian Foundation for Basic
Research (Grants No 17-02-00320 (VNR) and 18-02-00981
(YDF))

\end{document}